\newif\ifmnras
\mnrasfalse
\ifmnras
	\documentclass[useAMS,usenatbib]{mnras}
\else
	\documentclass[apj,numberedappendix]{emulateapj}
\fi
\usepackage{times}
\usepackage{graphics,epsf}
\usepackage{amsmath}                
\usepackage{amsfonts}               
\usepackage{amssymb}                
\usepackage{epsfig}                 
\usepackage{graphicx}
\usepackage{url}
\usepackage{tabularx}
\usepackage{booktabs}
\usepackage{hyperref}
\usepackage[dvipsnames]{xcolor}
\def \kms{\rm{km}$\rm{s}^{-1}$}

\def \cm{~\rm{cm}}
\def \s{~\rm{s}}

\def \km{~\rm{km}}
\def \kms{~\rm{km}~{\rm s}^{-1}}

\def \K{~\rm{K}}

\def \erg{~\rm{erg}}

\def \zams{\mathrm{ZAMS}}

\ifmnras
	\def \aap{A\&A}
	\def \aj{AJ}
	\def \apj{ApJ}
	\def \apjl{ApJ}
	\def \apjs{ApJS}
	\def \nat{Nature}
	\def \mnras{MNRAS}
\fi

\ifmnras
	\title[Pre-explosion dynamo in the cores of massive stars]{Pre-explosion dynamo in the cores of massive stars}

	\author[N. Soker \& A. Gilkis]{Noam Soker and Avishai Gilkis \\
	Department of Physics, Technion -- Israel, Institute of Technology, Haifa 32000, Israel;
	soker@physics.technion.ac.il;
	agilkis@technion.ac.il}
\fi

\begin{document}

\ifmnras
	\pagerange{\pageref{firstpage}--\pageref{lastpage}} \pubyear{2016}

	\maketitle

	\label{firstpage}
\else
	\title{Pre-explosion dynamo in the cores of massive stars}

	\author{Noam Soker}
	\author{Avishai Gilkis}
	\affil{Department of Physics, Technion -- Israel
	Institute of Technology, Haifa 32000, Israel;
	soker@physics.technion.ac.il; agilkis@technion.ac.il}
\fi

\begin{abstract}
We propose a speculative scenario where dynamo amplification of magnetic fields in the core convective shells of massive stars, tens of years to hours before they explode, leads to envelope expansion and enhanced mass loss rate, resulting in pre-explosion outbursts (PEOs).
The convective luminosity in the burning shells of carbon, neon, oxygen, and then silicon, are very high. Based on the behavior of active main sequence stars we speculate that the convective shells can trigger magnetic activity with a power of about 0.001 times the convective luminosity. Magnetic flux tubes might buoy outward, and deposit their energy in the outer parts of the envelope. This in turn might lead to the expansion of the envelope and to an enhanced mass loss rate. If a close binary companion is present, mass transfer might take place and lead to an energetic outburst. The magnetic activity requires minimum core rotation and that the stochastic magnetic activity be on its high phase. Only in rare cases these conditions are met, accounting for that only the minority of core collapse supernovae (CCSNe) experience PEO. Such a pre-explosion magnetic activity might have implications for the explosion mechanism itself.
\end{abstract}

\section{INTRODUCTION}
\label{sec:introduction}

The progenitors of some core collapse supernovae (CCSNe) experience one or more pre-explosion outbursts (PEOs) that are accompanied by high mass loss rate episodes. These might take place tens of years prior to explosion, to only days prior to explosion (e.g., \citealt{Foleyetal2007, Pastorelloetal2007, Smithetal2010, Mauerhanetal2013, Ofeketal2013, Pastorelloetal2013, Marguttietal2014, Ofeketal2014, SvirskiNakar2014, Marguttietal2016, Tartagliaetal2016}).
In some of these objects the PEO is a major outburst of a luminous blue variable (LBV), e.g., like the several outbursts in SN~2009ip \citep{Smithetal2010}.
In some cases the enhanced mass loss rate episode should occur as early as the core carbon-burning phase (e.g., \citealt{Moriyaetal2014, Marguttietal2016}).
Standard stellar evolution models do not account for PEOs.

There is an open question (e.g., \citealt{Levesqueetal2014, Marguttietal2016}) on whether the PEOs and the LBV major eruptions involve some kind of a binary interaction (e.g., \citealt{KashiSoker2010, Soker2013, McleySoker2014}), or whether these are driven by single-star processes (e.g., \citealt{Shaviv2000, Shaviv2001, Owockietal2004, Quataertetal2016, Moriya2014}).
Of course, there is the possibility that some PEOs are driven by binary interaction and some PEOs are driven by single star mechanisms.
Even models that are based on some binary interaction (e.g., \citealt{SokerKashi2013}) might require that the exploding star, the primary star, experiences an unstable phase prior to explosion. Such can be triggered by the last stages of the nuclear burning. Although most of the energy released in the burning is carried by neutrinos, and hence does not affect the stellar structure, there is a vigorous convection in the core.  Nuclear burning processes of carbon, neon, oxygen, and then silicon occur hundreds of years, for carbon burning, to hours, for silicon burning, before the explosion of CCSNe.

Several mechanisms for PEOs have been proposed.  First there is the fifty years old pair-instability mechanism \citep{RakavyShaviv1967, Barkatetal1967} that can lead to a huge mass loss before explosion (e.g., \citealt{Chenetal2014}). Another mechanism might result from radiation-driven instabilities (e.g., \citealt{BlaesSocrates2003}), that might take place in LBV stars (e.g.,  \citealt{Kiriakidisetal1993}).
\cite{SmithArnett2014} raise the possibility that hydrodynamic instabilities resulting from turbulent convection in the nuclear burning phases before explosion might cause enhanced mass loss, and/or might cause the star to swell and to interact with a binary companion (as suggested earlier by, e.g., \citealt{Soker2013}).

\cite{QuataertShiode2012} and \cite{ShiodeQuataert2014} propose that the convection in the core during the last stages of nuclear burning excites g-waves that can carry non-negligible amount of energy outward. The waves turn to p-waves in the envelope, and deposit most of their energy in the outer layers of the envelope. \cite{QuataertShiode2012} and \cite{ShiodeQuataert2014} further argue that although the waves carry a small fraction of the energy released by the nuclear burning, this amount of energy is sufficient to substantially increase the mass loss rate from the star.
The p-wave luminosity can be super-Eddington \citep{QuataertShiode2012}, and the energy dissipation leads to ejection of the outer layers of the envelope, or to expansion of the envelope  \citep{ShiodeQuataert2014}. \cite{ShiodeQuataert2014} estimate that $\approx 20 \%$ of the SN progenitors can excite $10^{46}-10^{48} \erg$ of energy in waves. These waves might drive mass loss within a few months to a decade of core collapse.

\cite{Soker2013} on the other hand suggests that the p-waves lead to a large and rapid envelope expansion rather than to a large mass loss episode. A binary companion then interacts with the expanded envelope. Mass accreted onto the secondary star can release further energy, and account for high velocity gas. \cite{McleySoker2014} further develop this scenario. Using the stellar evolutionary code MESA they show that energy deposited to the envelope is likely to lead to its expansion rather than to large mass ejection.

In general, the energy delivered to the envelope can act in several ways to enhance mass loss rate and to cause an outburst (e.g., \citealt{McleySoker2014}). The delivered energy can be dissipated in the envelope, e.g., if g-waves excited in the core are turned to sound waves that are dissipated in the envelope. This increases the luminosity of the star. Luminosity by itself can increase the mass loss rate by radiation pressure on ions, molecules or dust.
If convection cannot remove the extra thermal energy, the envelope expands. With a lower gravity on the surface, the mass loss rate is expected to be higher. Another effect is the direct pressure increase in the envelope caused by expanding waves (wave pressure). This can lead to a rapid expansion of the envelope, again leading to enhanced mass loss rate. Finally, in the case of the presence of a close stellar companion, the expansion of the envelope might lead to a strong and violent binary interaction. The companion might form a common envelope, and/or it can accrete mass and release energy. In both cases large amounts of gravitational energy can lead to a strong outburst with a substantially enhanced mass loss rate.

The model for PEOs must involve an ingredient that is relatively rare, about one in ten CCSNe (e.g., \citealt{Marguttietal2016}). We know that at least the progenitor of SN~1987A, for example, did not experience such a phase within years before explosion.
In the present study we speculate that magnetic activity in the core has the required properties of delivering energy from the core to the envelope, and be limited to a small fraction of all CCSNe.
In section \ref{sec:dynamo} we outline a phenomenological scenario for dynamo activity. Based on the behavior of main sequence stars we speculate that strong magnetic activity requires core rotation and even then might be stochastic. This explains the rare occurrence of such cases. In section \ref{sec:stellar} we present pre-explosion stellar models emphasizing the relevant convective shells.
In section \ref{sec:tubes} we follow the evolution of magnetic flux tubes, and the way they can influence the envelope. We summarize in section \ref{sec:summary}

\section{CORE DYNAMO}
\label{sec:dynamo}

The development of a dynamo model for the core of pre-exploding stars is beyond the scope of this paper. We simply infer from main sequence stars, and make the following assumptions.

(1) \emph{Dynamo in convective shells.} In low mass main sequence stars, such as the sun, magnetic field amplification takes place in the convective envelope, mainly in the lower boundary of the convective zone. Magnetic field lines are concentrated there (e.g., \citealt{KitchatinovNepomnyashchikh2016}) and lead to the formation of magnetic flux tubes. We assume that magnetic activity takes place at the bottom of convective shells in the core.

(2) \emph{Magnetic activity power.} In active main sequence stars the X-ray luminosity reaches a value of $L_X \simeq 0.001 L_{\rm bol}$, where $L_{\rm bol}$ is the bolometric luminosity of the star (e.g., \citealt{Pizzolatoetal2003}). The X-ray emission results from magnetic activity above the stellar photosphere. In such stars the convection carries the entire energy in the outer part of the envelope, such that the convective luminosity (see section \ref{sec:stellar}) is $L_{\rm conv} \approx L_{\rm bol} \approx 10^3 L_x \approx 10^3 L_B$.
The X-ray power is about equal to the magnetic power of the star, $L_X \approx L_B$, that in turn results from the convection.
We assume that the magnetic activity in convective shells in the core of pre-exploding stars obeys a similar crude ratio between $L_B$ and $L_{\rm conv}$, and can reach a power of up to $L_B \approx 0.001 L_{\rm conv}$. This condition is met only in a small fraction of all stars, probably those where the core has some minimum value of angular velocity.
As \cite{ShiodeQuataert2014} have $L_{\rm wave} \approx 0.01 L_{\rm conv}$, under the assumptions we make here, the magnetic power of the core can reach a value of $L_B \approx 0.1 L_{\rm wave}$.

(3) \emph{Buoyancy of magnetic flux tubes.} We assume that magnetic flux tubes buoy to the envelope. This is treated in section \ref{sec:tubes}.

(4) \emph{Envelope expansion.} Upon deposition of sufficient energy to the envelope by the magnetic flux tubes, the star expands and interacts with a binary companion, or the mass loss rate simply increases \citep{HarpazSoker2009}. The envelope expansion is caused by the pressure inside the magnetic flux tubes (they simply occupy a volume), and by dissipation of magnetic energy that increases the thermal energy of the envelope. Mass loss then increases either by the higher luminosity and larger radius of the star (hence lower gravity), or by an interaction with a close companion (if present) that can deposit more energy by spiraling-in or by accreting mass from the envelope.

The assumption that the dynamo in pre-explosion massive stars is similar to that of main sequence star is a very strong assumption. Moreover, we assume that the dynamo in the core of massive stars behaves as the dynamo in the envelope of main sequence stars. This strong assumption  will have to be examined in future studies. We do not have the tools to examine this assumption. At this point we make this assumption based on a general view that there is something generic and universal in the operation of dynamos in rotating-convective (or turbulent) bodies, from stars to accretion disks, to galaxies (e.g. \citealt{BlackmanField2000, VishniacCho2001}).
Note that since in massive stars the dynamo activity takes place in the core, the manifestation of the energy will not be in magnetic flares and X-ray emission as in main sequence stars, but rather we expect that the energy be deposited inside the envelope. Therefore, the observed low X-ray luminosity of LBV stars (e.g. \citealt{Nazeetal2012}), does not contradict our assumption of strong dynamo activity in rapidly rotating cores.

\cite{HarpazSoker2009} suggest that LBV major eruptions, such as the 1837-1856 Great Eruption of
$\eta$ Carinae, might be triggered by magnetic activity cycles. They suggest that a strong magnetic field region can be built in the radiative zone below an outer convective region and above the  convective core of an evolved LBV massive star. This magnetic energy, they speculate, might trigger major eruptions of some LBV stars. The eruption will be particularly strong if the star interacts with a binary companion, as in the Great Eruption of $\eta$ Carinae. We here consider more evolved stars, with more vigorous core convection, and the triggering of PEOs. In cases of eruptive mass loss events, our scenario requires a companion to interact with the expanded envelope.
The presently proposed scenario, therefore, connects PEOs to major LBV outbursts that occur a century or more before eruptions, such as the Great Eruption of $\eta$ Carinae; both types are proposed to be triggered by magnetic activity.

\section{STELLAR MODELS}
\label{sec:stellar}

\begin{figure}
\begin{tabular}{c}
\ifmnras
	{\includegraphics*[scale=0.5]{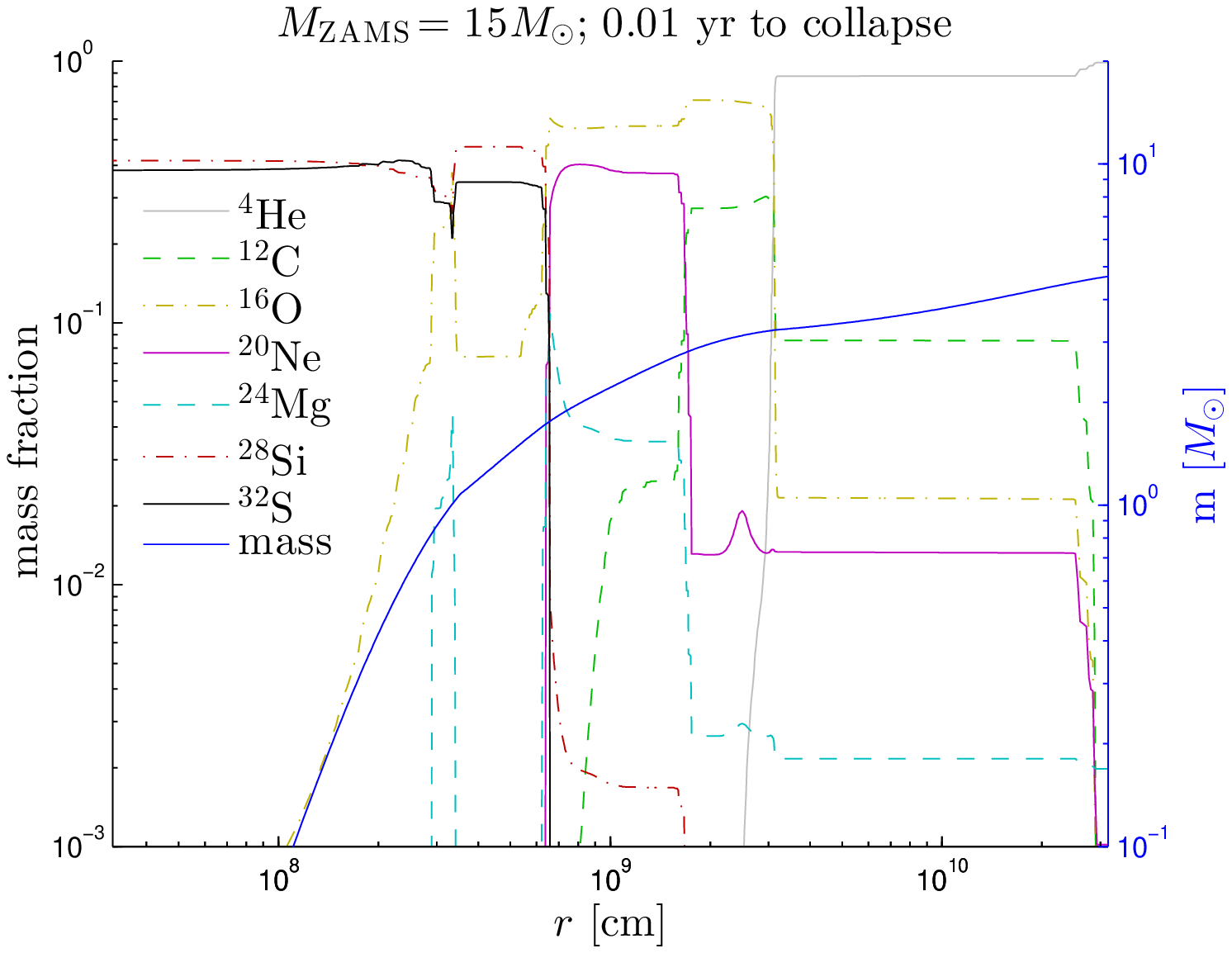}} \\
	{\includegraphics*[scale=0.5]{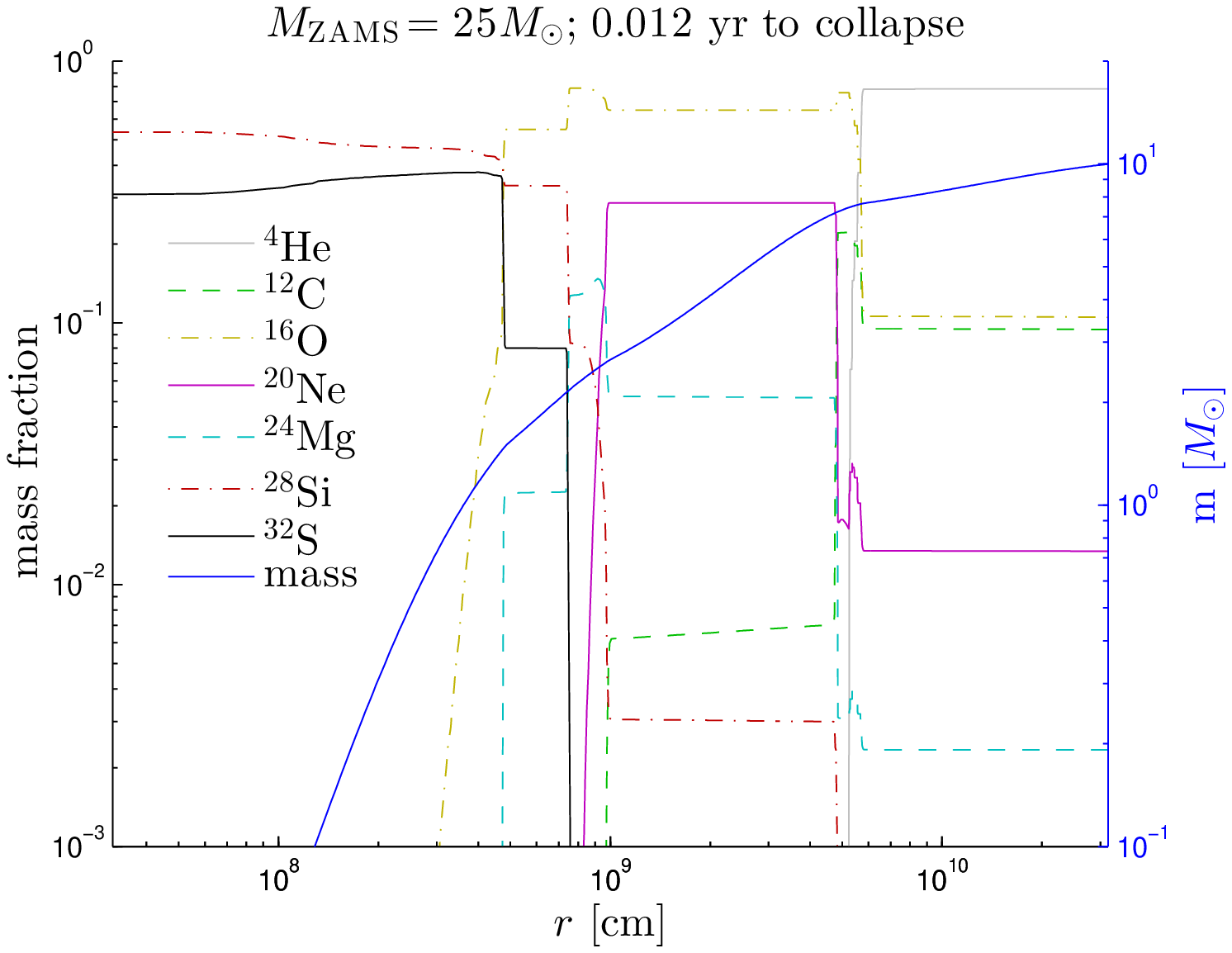}} \\
\else
	{\includegraphics*[scale=0.52]{M15composition.eps}} \\
	{\includegraphics*[scale=0.52]{M25composition.eps}} \\
\fi
\end{tabular}
      \caption{Detailed core composition at late stages of oxygen shell burning,
      for our $M_\zams = 15 M_\odot$ model (top) and $M_\zams = 25 M_\odot$ model (bottom).
      The stellar mass, stellar radius, effective temperature, and luminosity, of the two models at the times shown just before collapse are
      $(M, R, T_{\rm eff}, L)_{\rm collpase}=
      (12.3 M_\odot, 1065 R_\odot, 3167 \K, 1.03 \times 10^5 L_\odot)$
          for the $M_\zams = 15 M_\odot$ model,  and
      $(M, R, T_{\rm eff}, L)_{\rm collpase}=
      (11.1 M_\odot, 500 R_\odot, 5919 \K, 2.76 \times 10^5 L_\odot)$
        for the $M_\zams = 25 M_\odot$ model.
      The solid blue lines show the enclosed mass as function of radial coordinate.
      The region shown extends from $r=10^{7.5}\cm$ to $r=10^{10.5}\cm$.}
      \label{fig:composition}
\end{figure}

\begin{figure}
\begin{tabular}{c}
\ifmnras
	{\includegraphics*[scale=0.5]{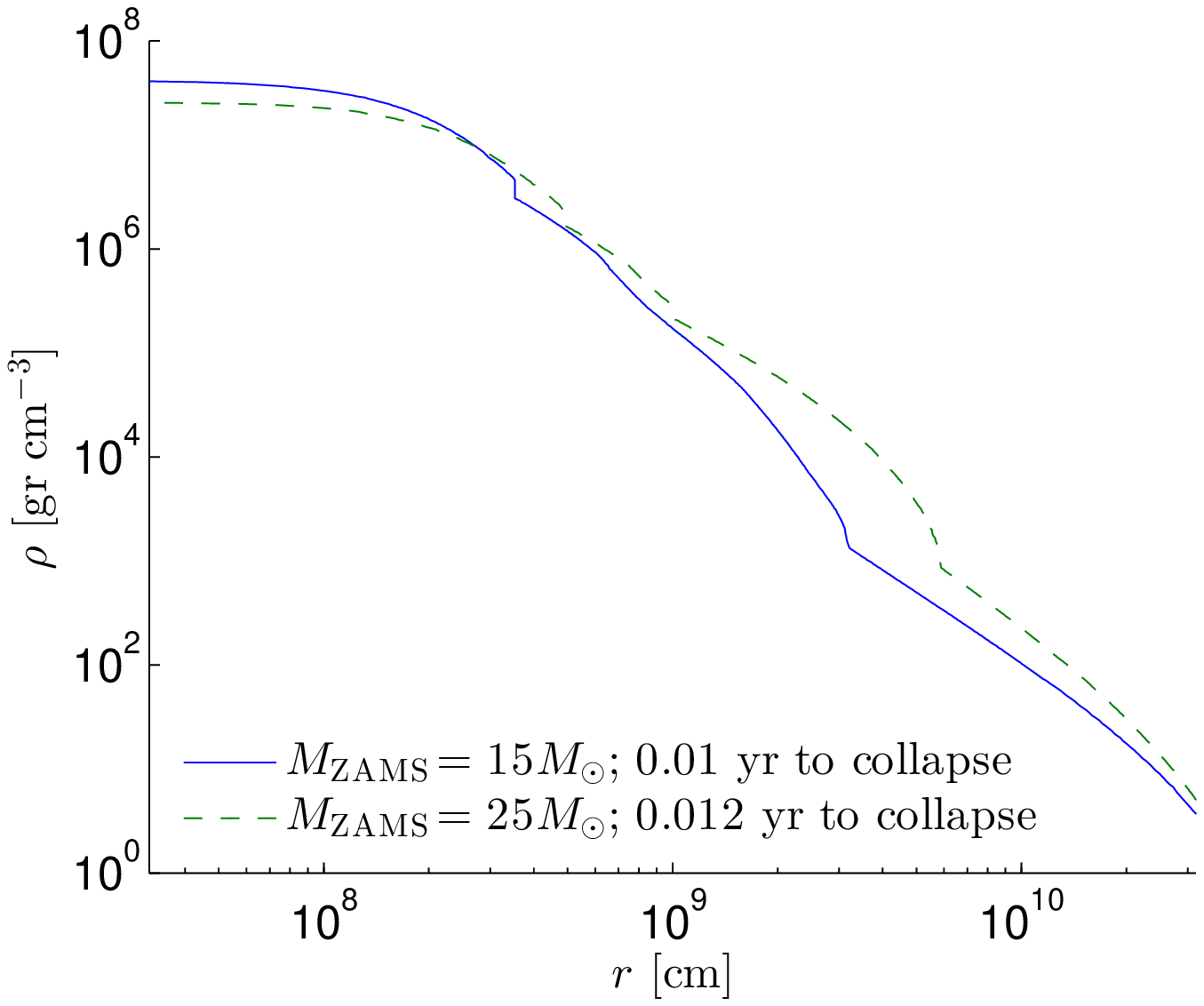}} \\
	{\includegraphics*[scale=0.5]{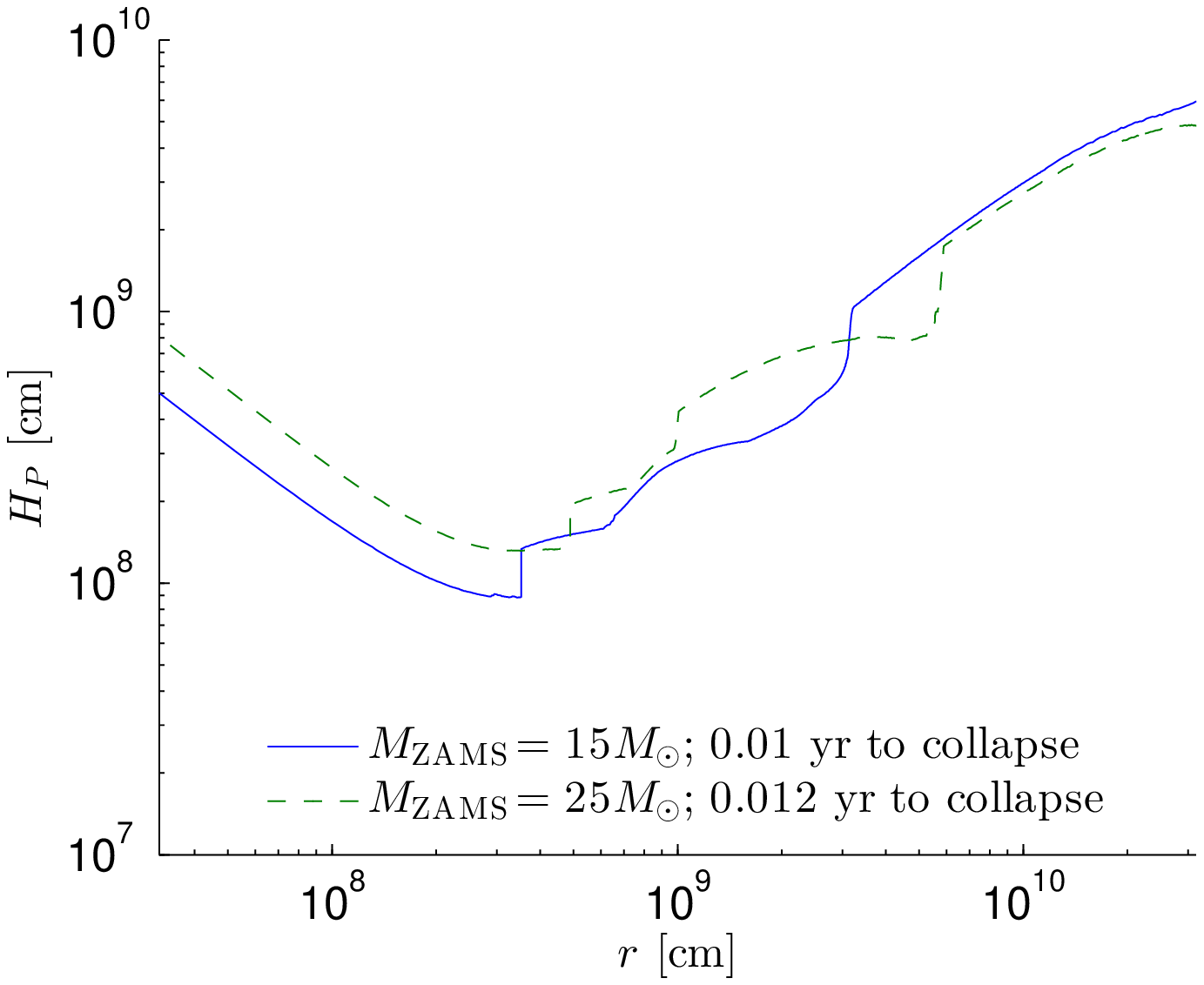}} \\
\else
	{\includegraphics*[scale=0.52]{density.eps}} \\
	{\includegraphics*[scale=0.52]{height.eps}} \\
\fi
\end{tabular}
      \caption{\textit{Top:}
      Density profile in the region between
      $r=10^{7.5}\cm$ to $r=10^{10.5}\cm$
      for our two stellar models at late stages of oxygen shell burning.
      \textit{Bottom:}
      The pressure scale height for our stellar models.}
      \label{fig:density}
\end{figure}

We evolve two stellar models to derive some quantities to be used in the next section,
using Modules for Experiments in Stellar Astrophysics (MESA version 7624; \citealt{Paxton2011,Paxton2013,Paxton2015}).
Our models have a metallicity of $Z=0.014$,
and zero-age main sequence (ZAMS) masses of $M_\zams = 15 M_\odot$ and $M_\zams = 25 M_\odot$.
The initial luminosity and effective surface temperature are $L_\zams = 1.9 \times 10^4 L_\odot$ and $ T_\zams = 3.1 \times 10^4 \K$, respectively, for the $M_\zams = 15 M_\odot$ model, and  $L_\zams = 7.5 \times 10^4 L_\odot$ and $ T_\zams = 3.8 \times 10^4 \K$ for the $M_\zams = 25 M_\odot$ model.
Convection is treated according to the Mixing-Length Theory with $\alpha_\mathrm{MLT}=1.5$.
Rotation is treated using the `shellular approximation' \citep{Meynet1997},
which assumes a constant angular velocity $\Omega$ in isobaric shells.
The initial rotation is set as $0.1$ of the breakup rotation,
which corresponds to an equatorial surface velocity of $v_\zams \approx 70 \kms$.
Mass loss during the main sequence phase is treated according to the results of \cite{Vink2001}. During the giant phase,
the mass loss depends on the surface luminosity and temperature according to the fit of \cite{deJager1988}. The mass loss rate is enhanced by rotation (e.g., \citealt{Heger2000,Maeder2000}).
The resulting pre-collapse masses are $M_\mathrm{collapse} = 12.3 M_\odot$
for the $M_\zams = 15 M_\odot$ model,
and $M_\mathrm{collapse} = 11.1 M_\odot$
for the $M_\zams = 25 M_\odot$ model.
While the model with an initially larger mass is lighter at late evolutionary stages (because of intensive mass loss),
it has a helium core mass of $M_\mathrm{core}=10.1 M_\odot$
compared to $M_\mathrm{core}=4.9 M_\odot$ for the $M_\zams = 15 M_\odot$ model.
Rotationally-driven mixing and angular momentum transport are accounted for, as described by \cite{Paxton2013}.
The Spruit-Tayler dynamo \citep{Spruit2002} is included in our models,
effectively transporting angular momentum from the core outward,
resulting in late-stage core rotation periods of hundreds to thousands of seconds.
The detailed compositions of our models at oxygen shell burning are presented in Figure \ref{fig:composition},
and the corresponding pressure scale height and density profiles are shown in Figure \ref{fig:density}.

\begin{figure*}
\begin{tabular}{ccc}
\ifmnras
	{\includegraphics*[scale=0.34]{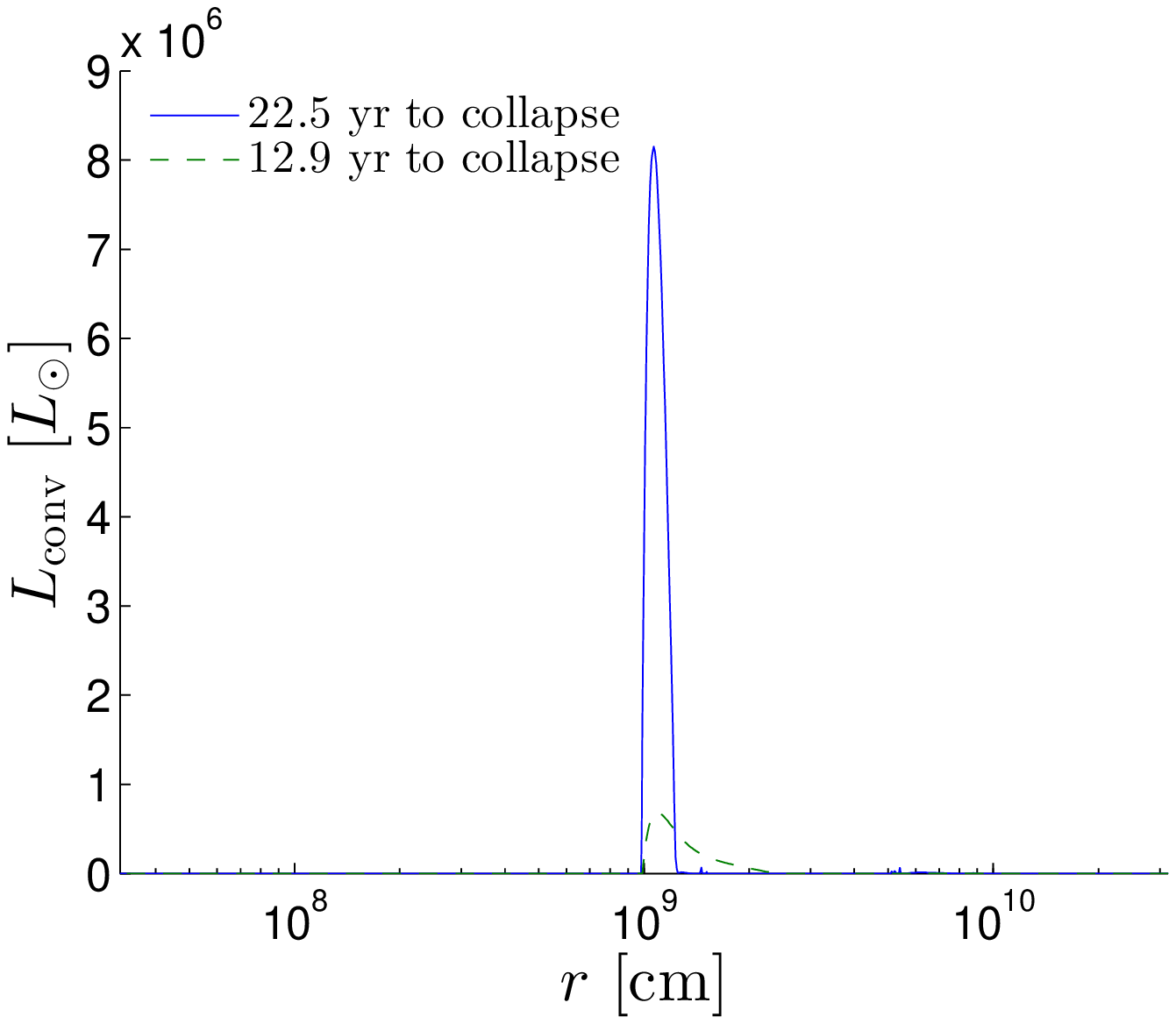}} &
	{\includegraphics*[scale=0.34]{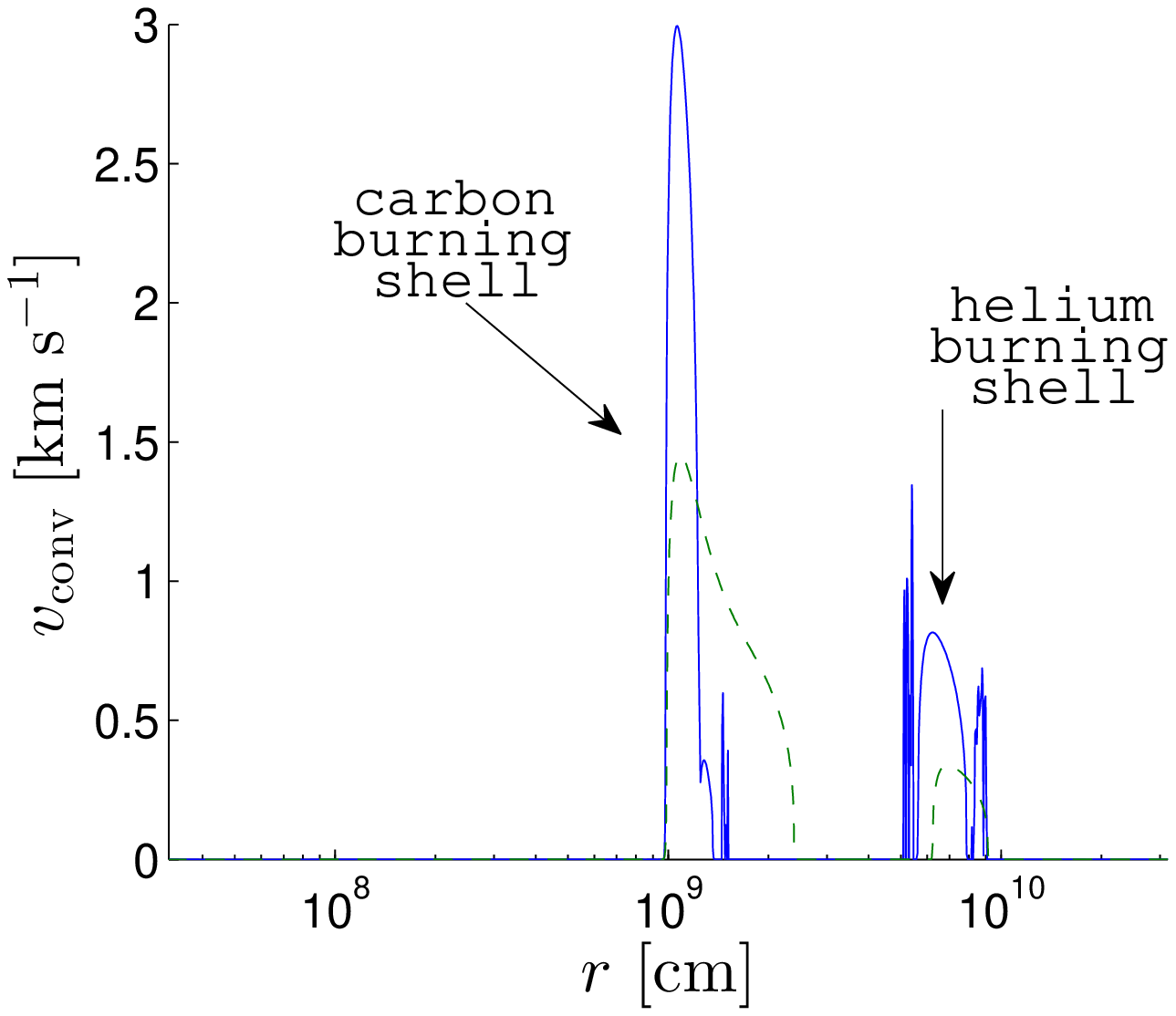}} &
	{\includegraphics*[scale=0.34]{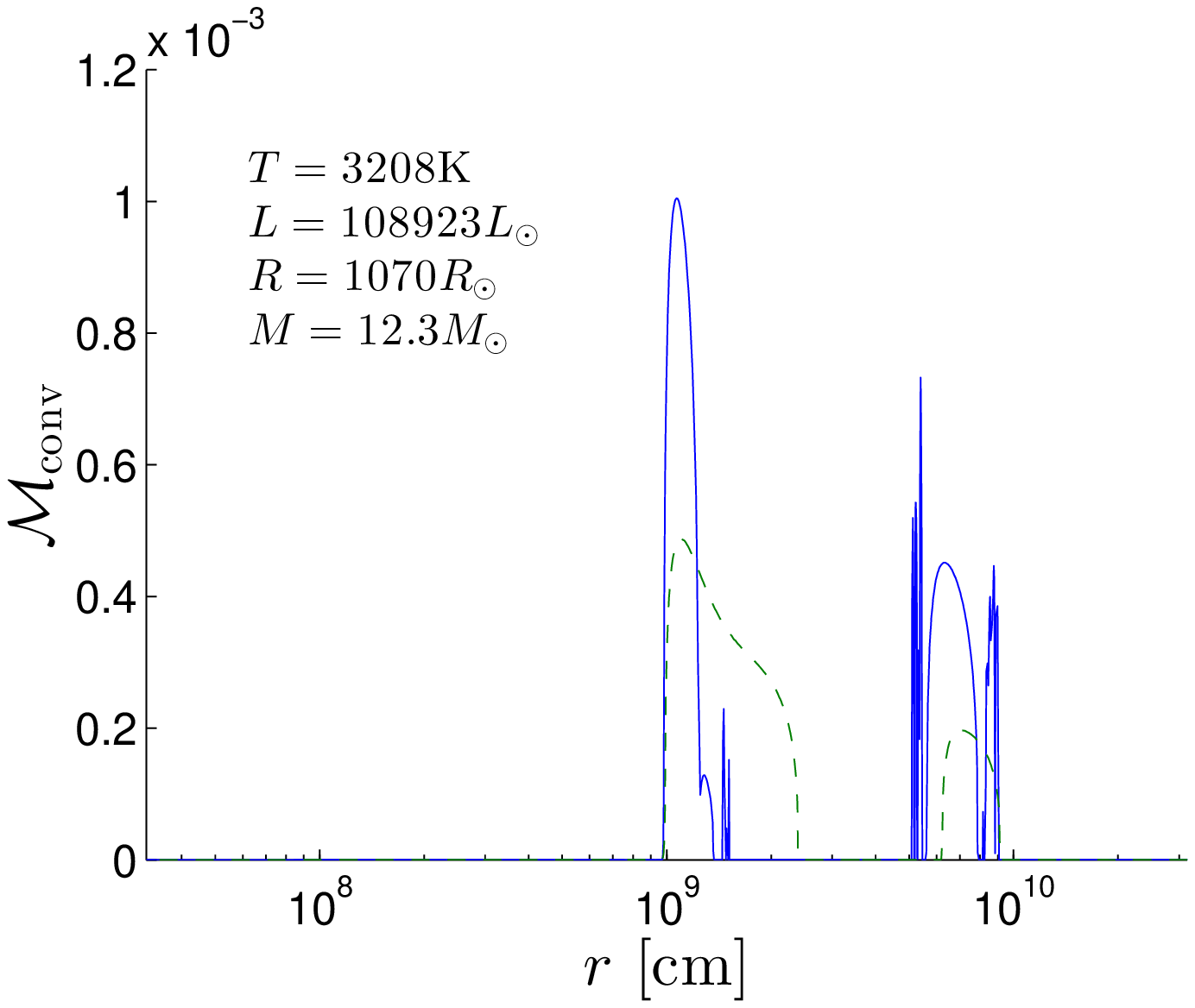}} \\
	{\includegraphics*[scale=0.34]{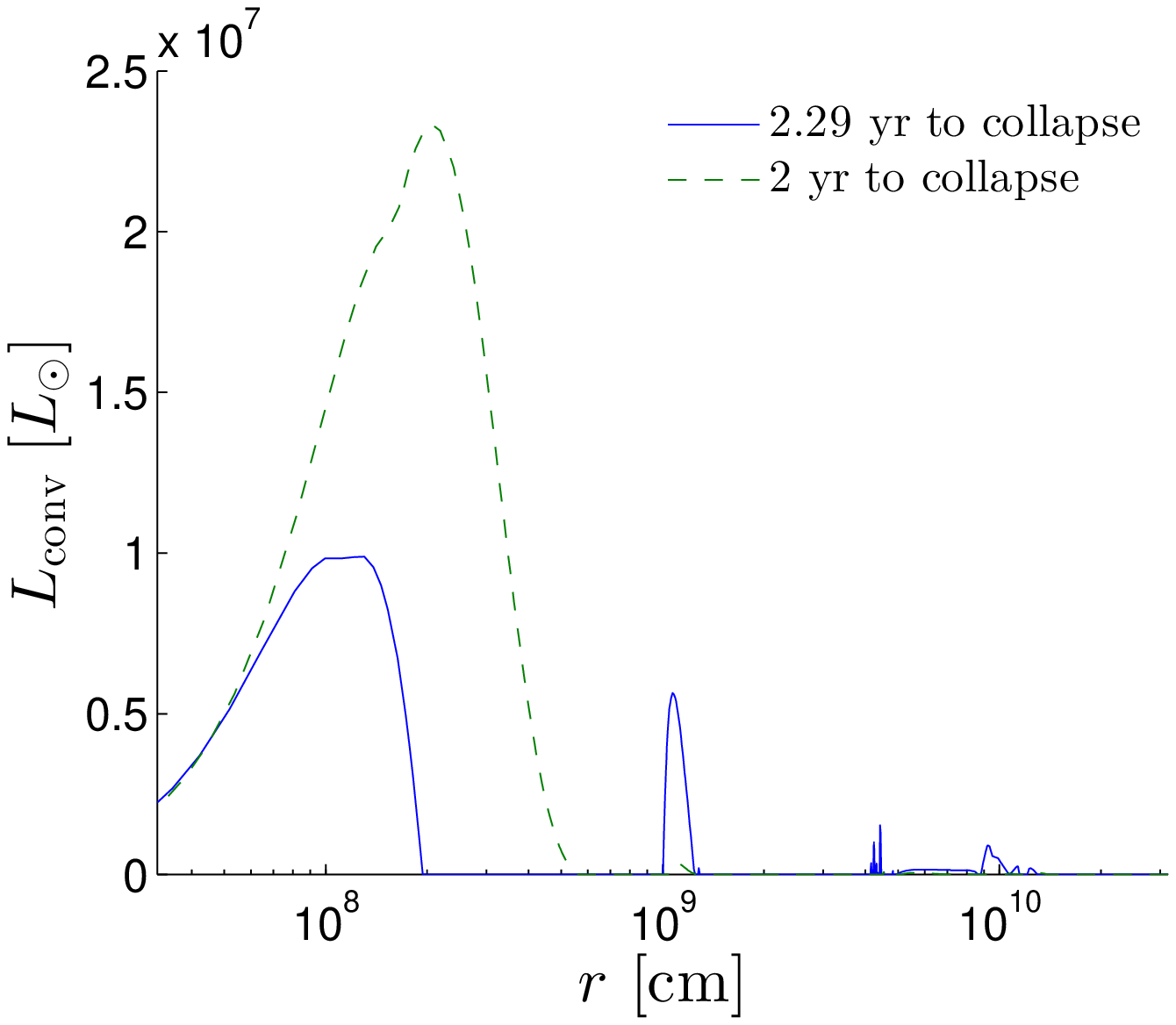}} &
	{\includegraphics*[scale=0.34]{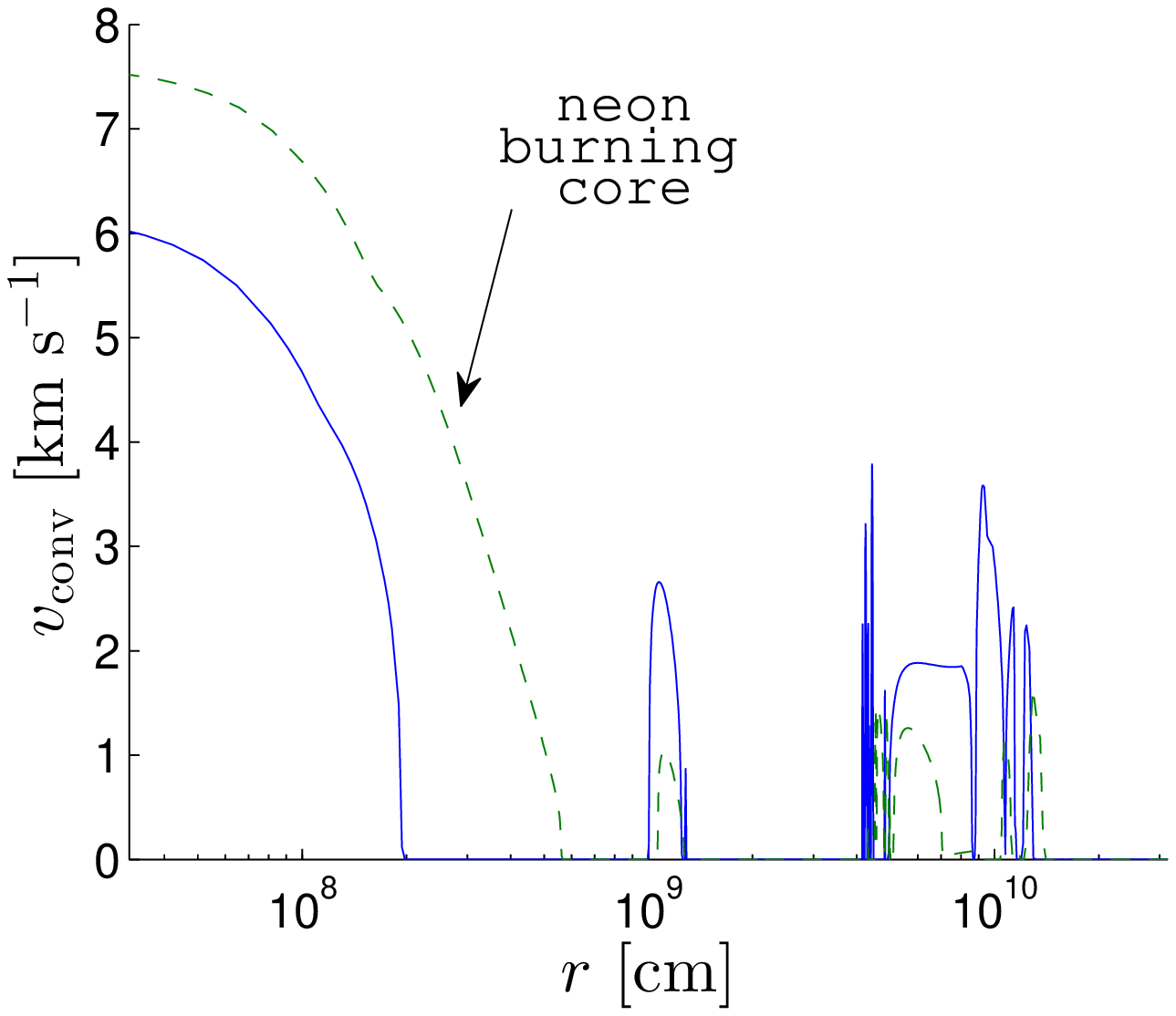}} &
	{\includegraphics*[scale=0.34]{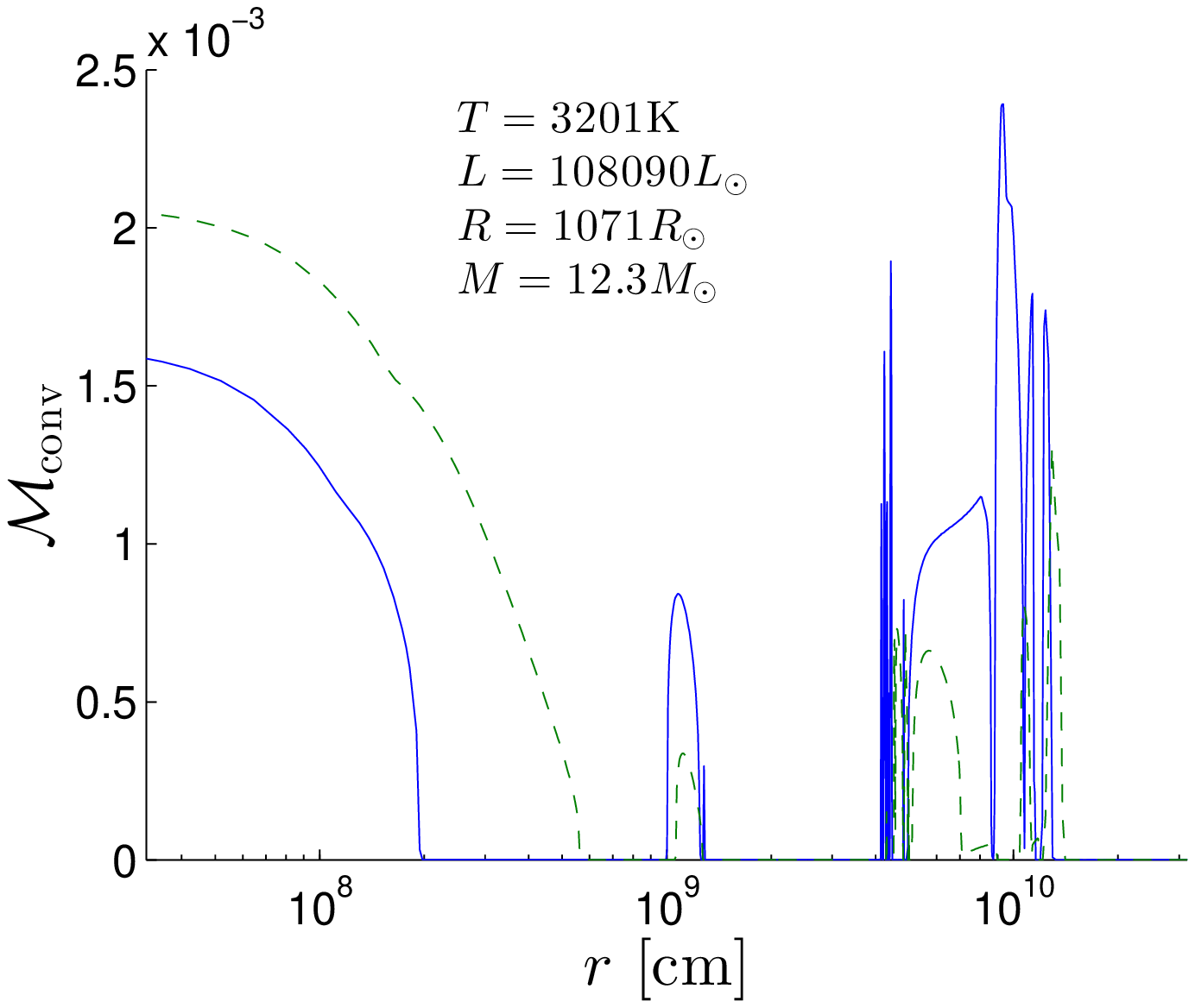}} \\
	{\includegraphics*[scale=0.34]{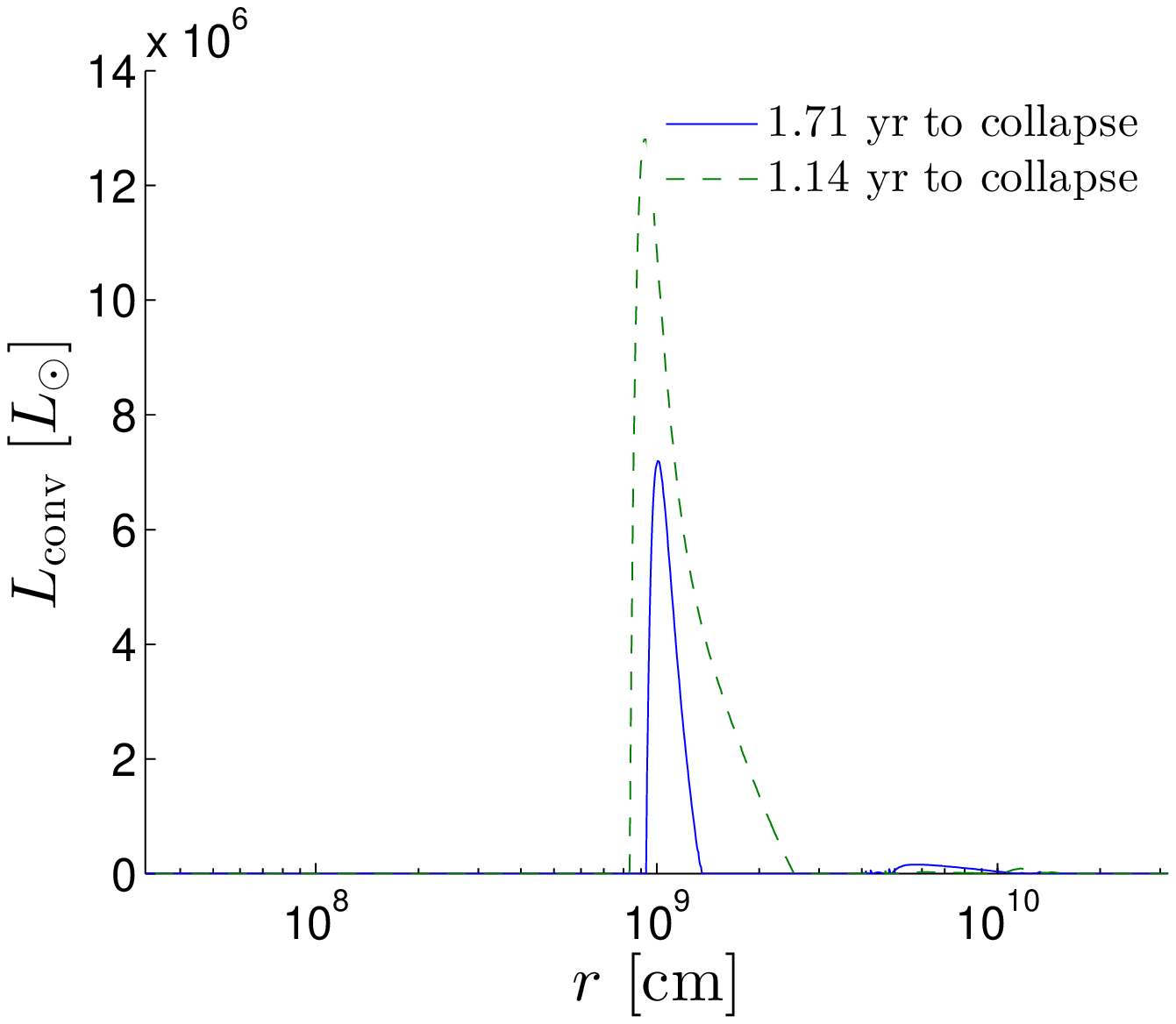}} &
	{\includegraphics*[scale=0.34]{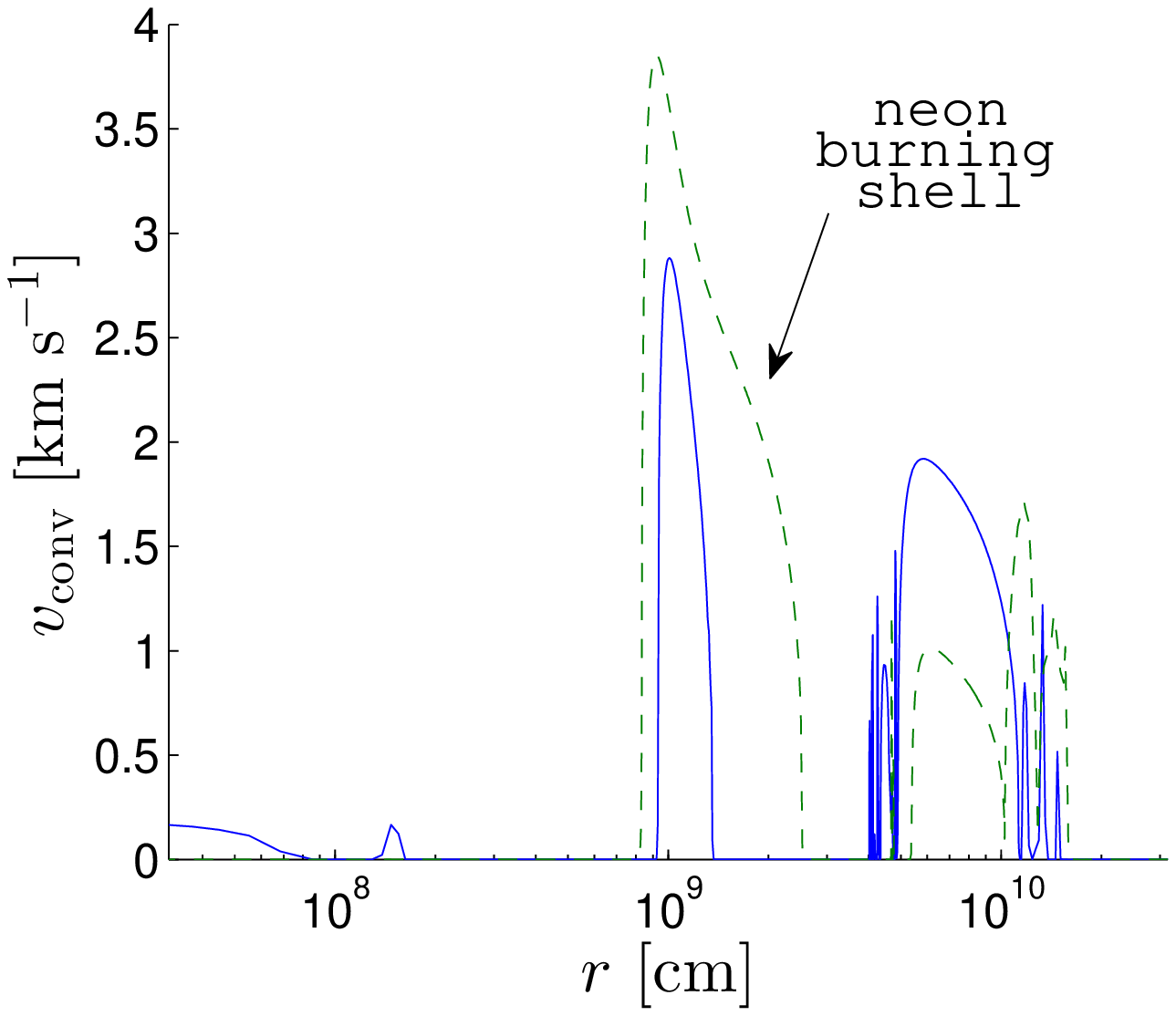}} &
	{\includegraphics*[scale=0.34]{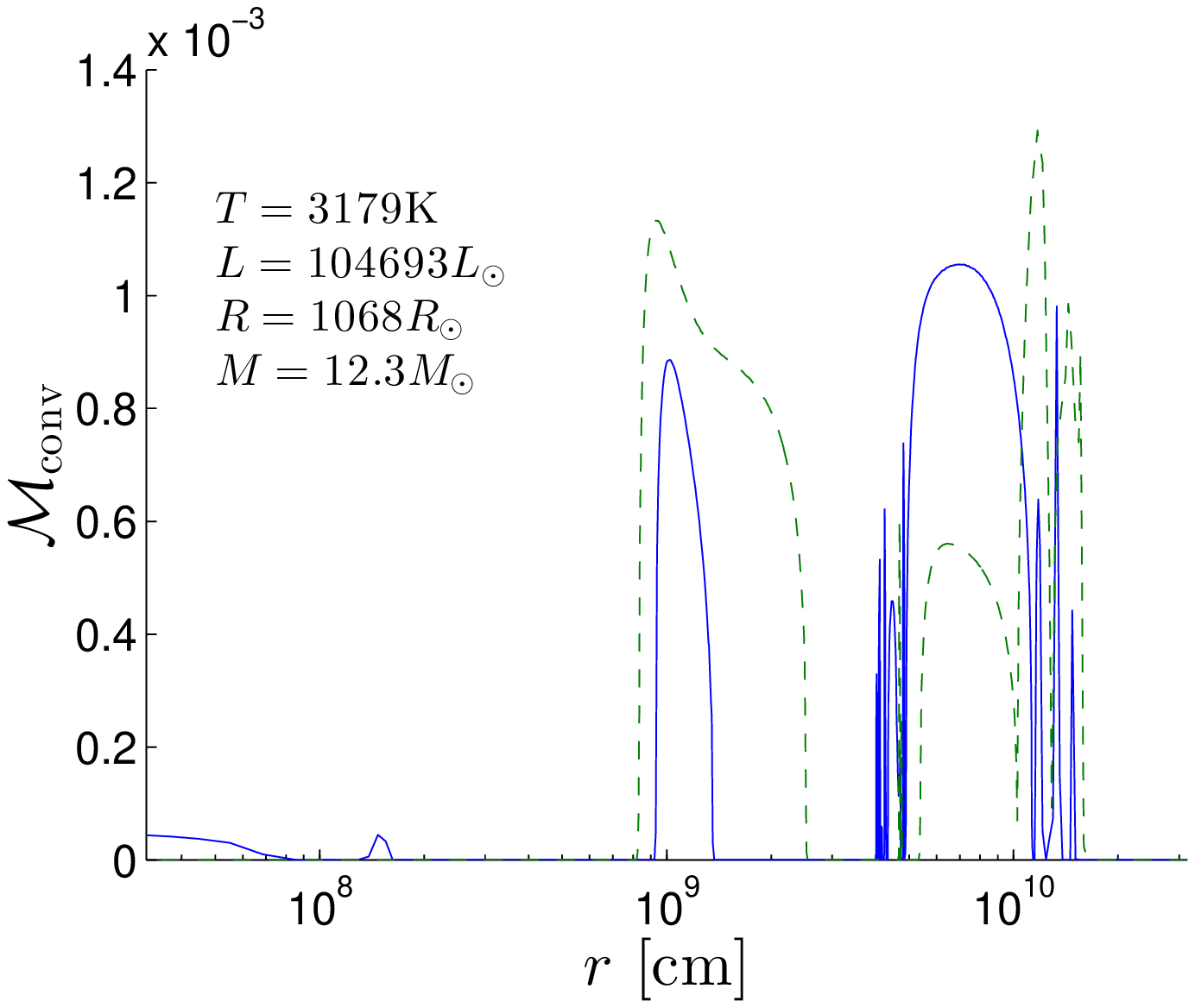}} \\
	{\includegraphics*[scale=0.34]{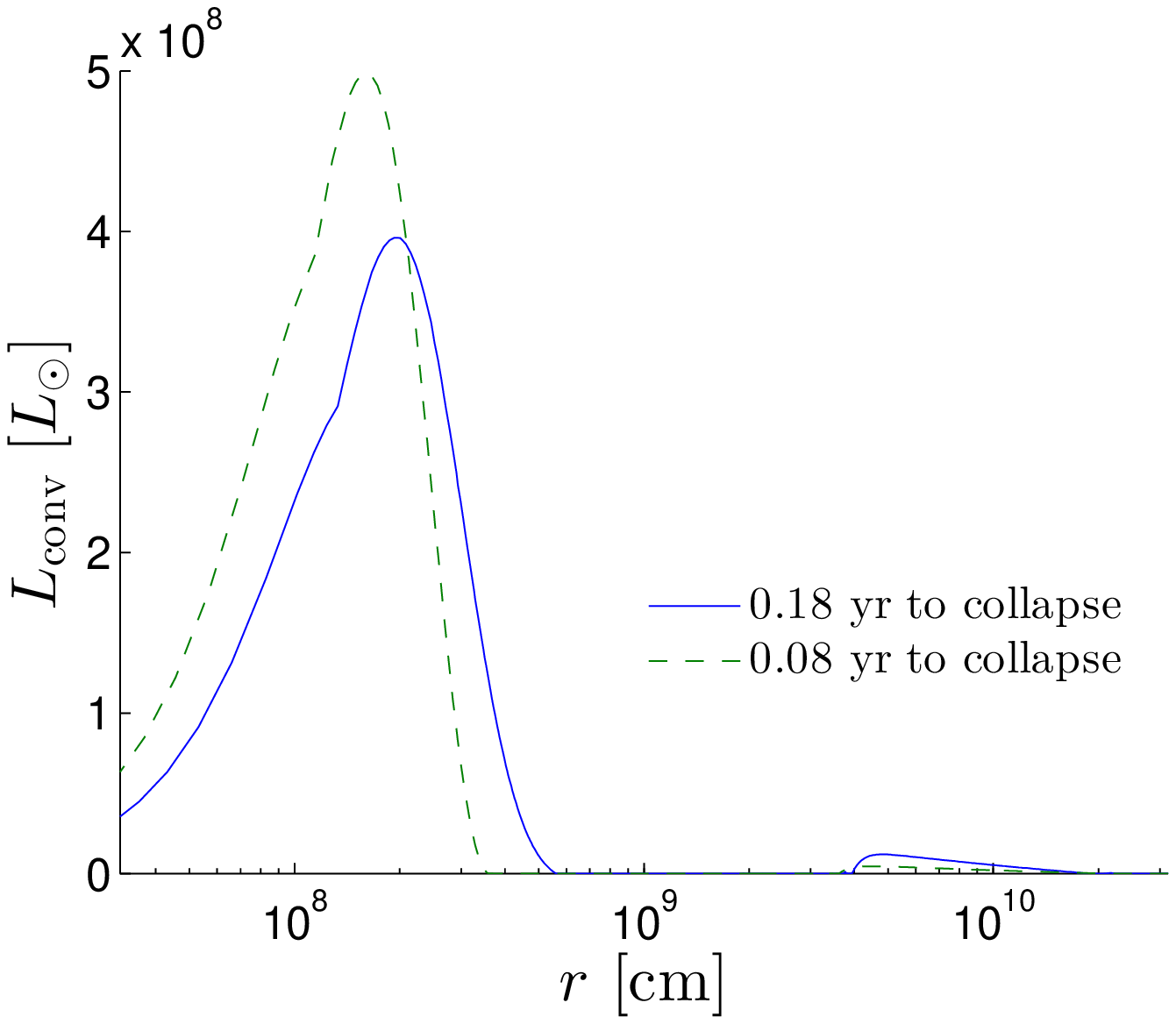}} &
	{\includegraphics*[scale=0.34]{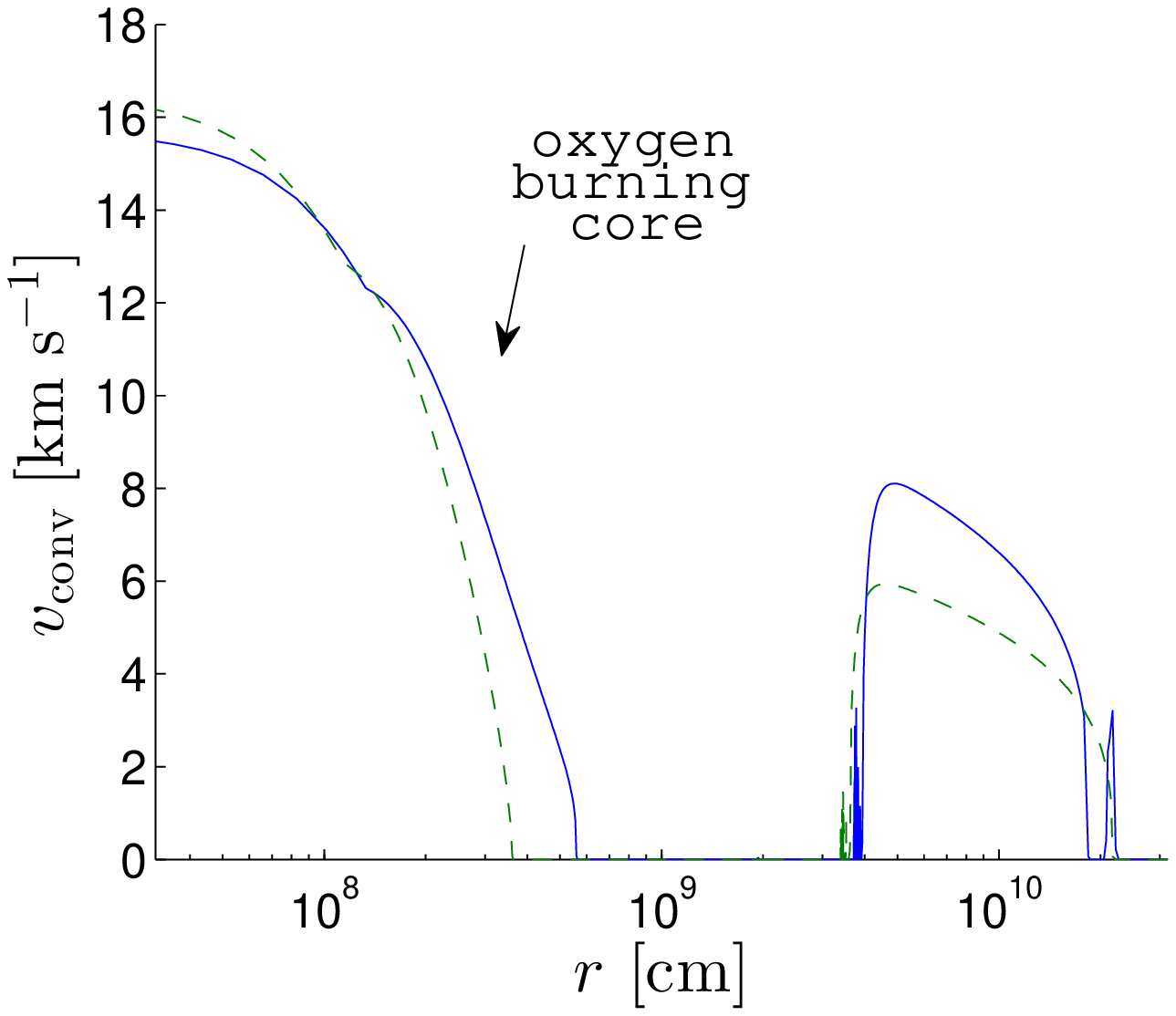}} &
	{\includegraphics*[scale=0.34]{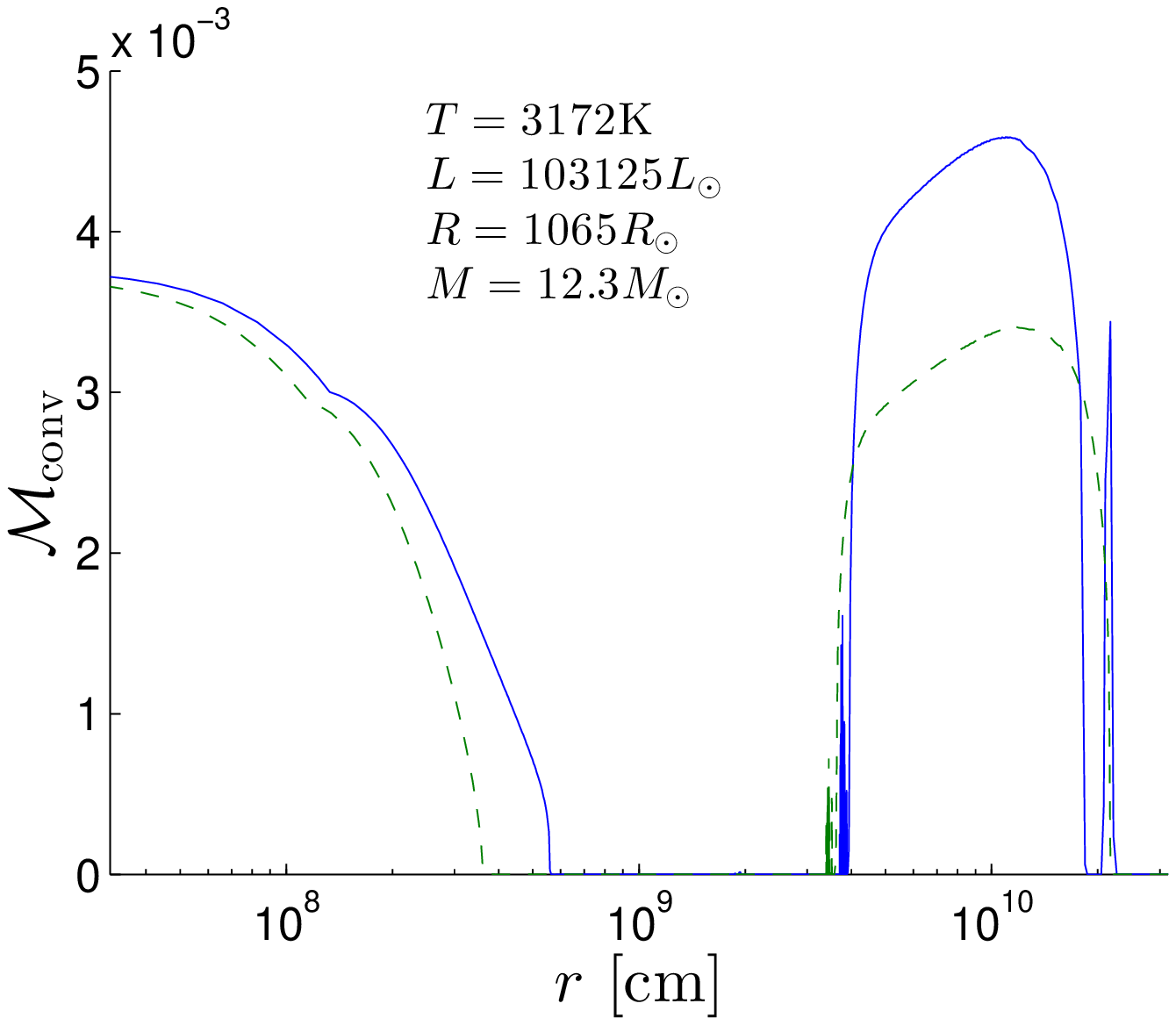}} \\
	{\includegraphics*[scale=0.34]{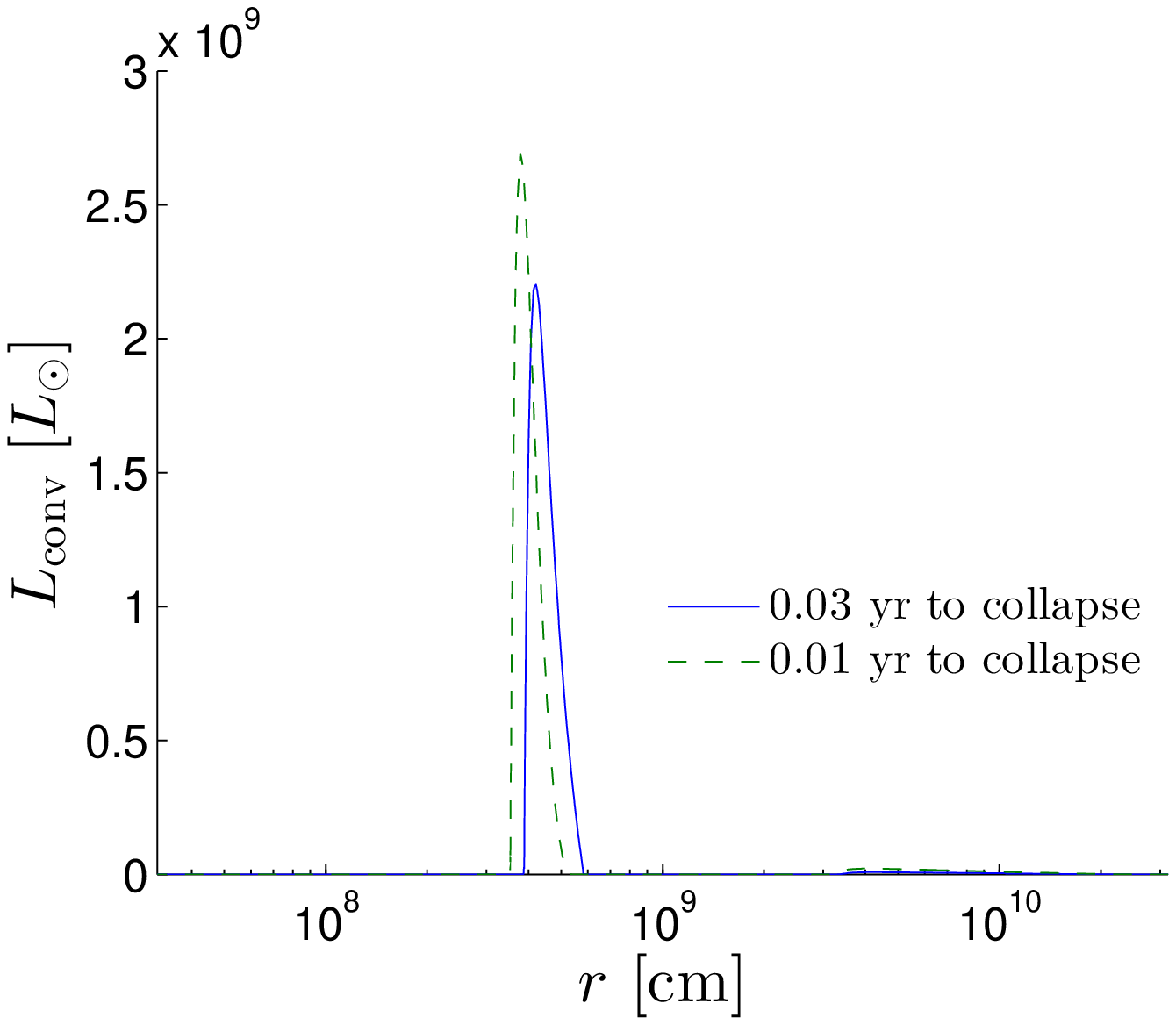}} &
	{\includegraphics*[scale=0.34]{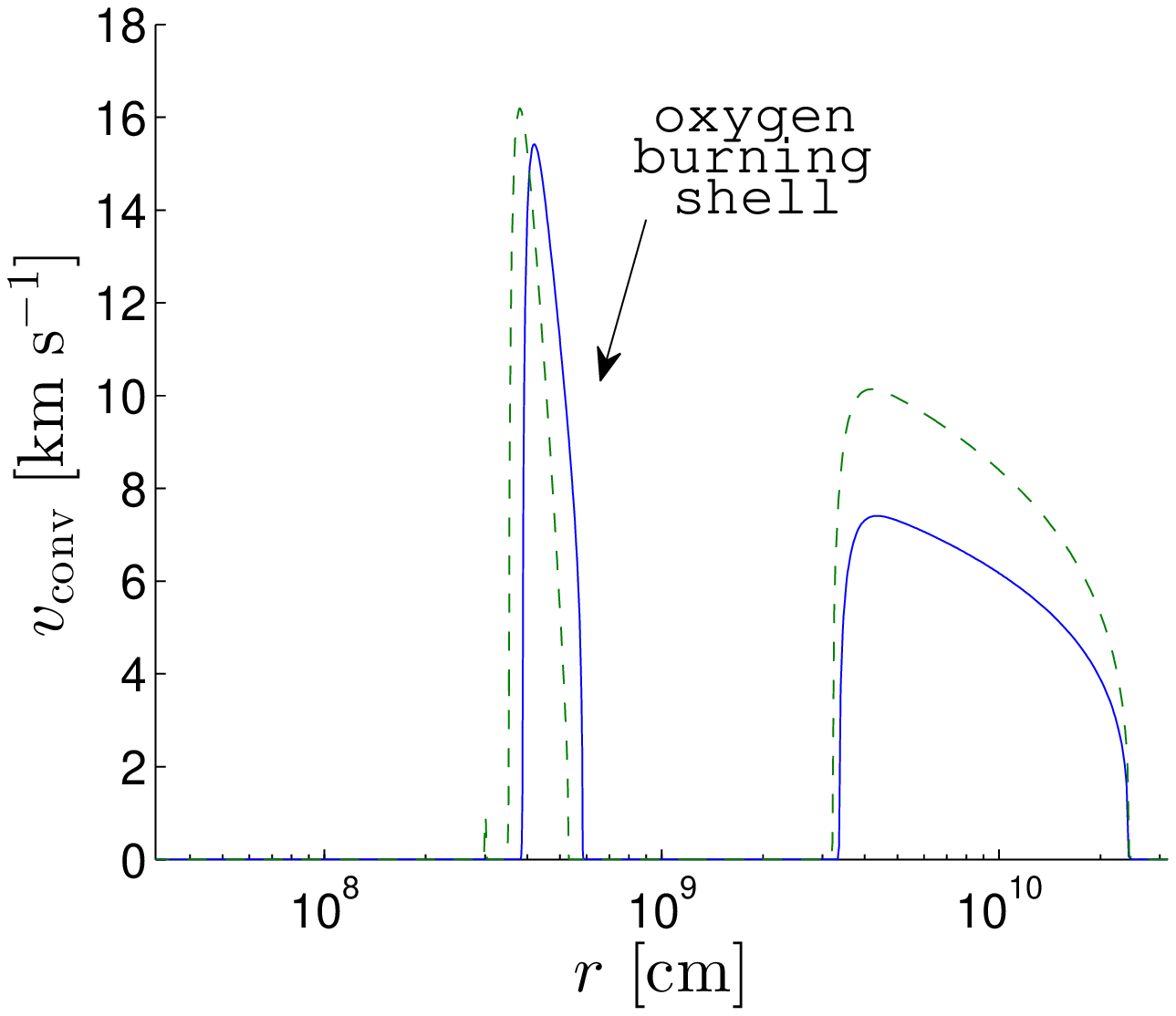}} &
	{\includegraphics*[scale=0.34]{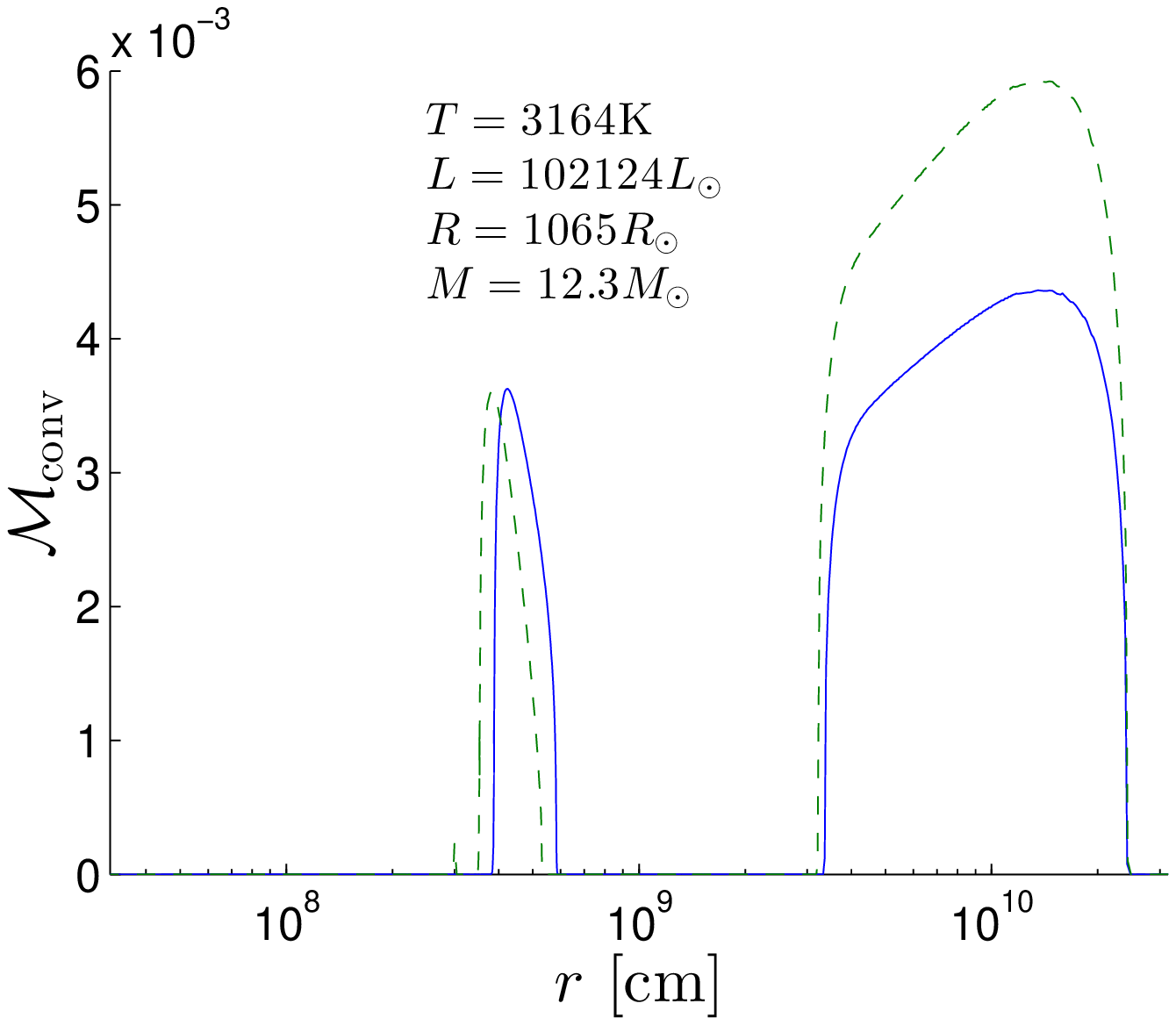}} \\
\else
	{\includegraphics*[scale=0.35]{M15LumCShell.eps}} &
	{\includegraphics*[scale=0.35]{M15VelCShell.eps}} &
	{\includegraphics*[scale=0.35]{M15MachCShell.eps}} \\
	{\includegraphics*[scale=0.35]{M15LumNeCore.eps}} &
	{\includegraphics*[scale=0.35]{M15VelNeCore.eps}} &
	{\includegraphics*[scale=0.35]{M15MachNeCore.eps}} \\
	{\includegraphics*[scale=0.35]{M15LumNeShell.eps}} &
	{\includegraphics*[scale=0.35]{M15VelNeShell.eps}} &
	{\includegraphics*[scale=0.35]{M15MachNeShell.eps}} \\
	{\includegraphics*[scale=0.35]{M15LumOCore.eps}} &
	{\includegraphics*[scale=0.35]{M15VelOCore.eps}} &
	{\includegraphics*[scale=0.35]{M15MachOCore.eps}} \\
	{\includegraphics*[scale=0.35]{M15LumOShell.eps}} &
	{\includegraphics*[scale=0.35]{M15VelOShell.eps}} &
	{\includegraphics*[scale=0.35]{M15MachOShell.eps}} \\
\fi
\end{tabular}
      \caption{Properties of the convective zones at five evolutionary phases for the stellar model with a ZAMS mass of $M_\zams = 15 M_\odot$. At each phase two times are compared. From top to bottom:
      Carbon shell burning, neon core burning, neon shell burning, oxygen core burning, and oxygen shell burning
      (carbon burning and neon burning occur simultaneously in shells, and we define these evolutionary stages according to the preceding core burning).
      The presented quantities are, from left to right, convective luminosity (eq. \ref{eq:ConvLum}), convective velocity, and convective Mach number.
      The time prior to the core-collapse is indicated for the two times in the left-most panel of each row. The stellar mass, radius, luminosity, and effective temperature
      are given in the right-most panel of each row for the later of the two times.
      The region shown extends from $r=10^{7.5}\cm$ to $r=10^{10.5}\cm$.}
      \label{fig:M15Conv}
\end{figure*}

\begin{figure*}
\begin{tabular}{ccc}
\ifmnras
	{\includegraphics*[scale=0.34]{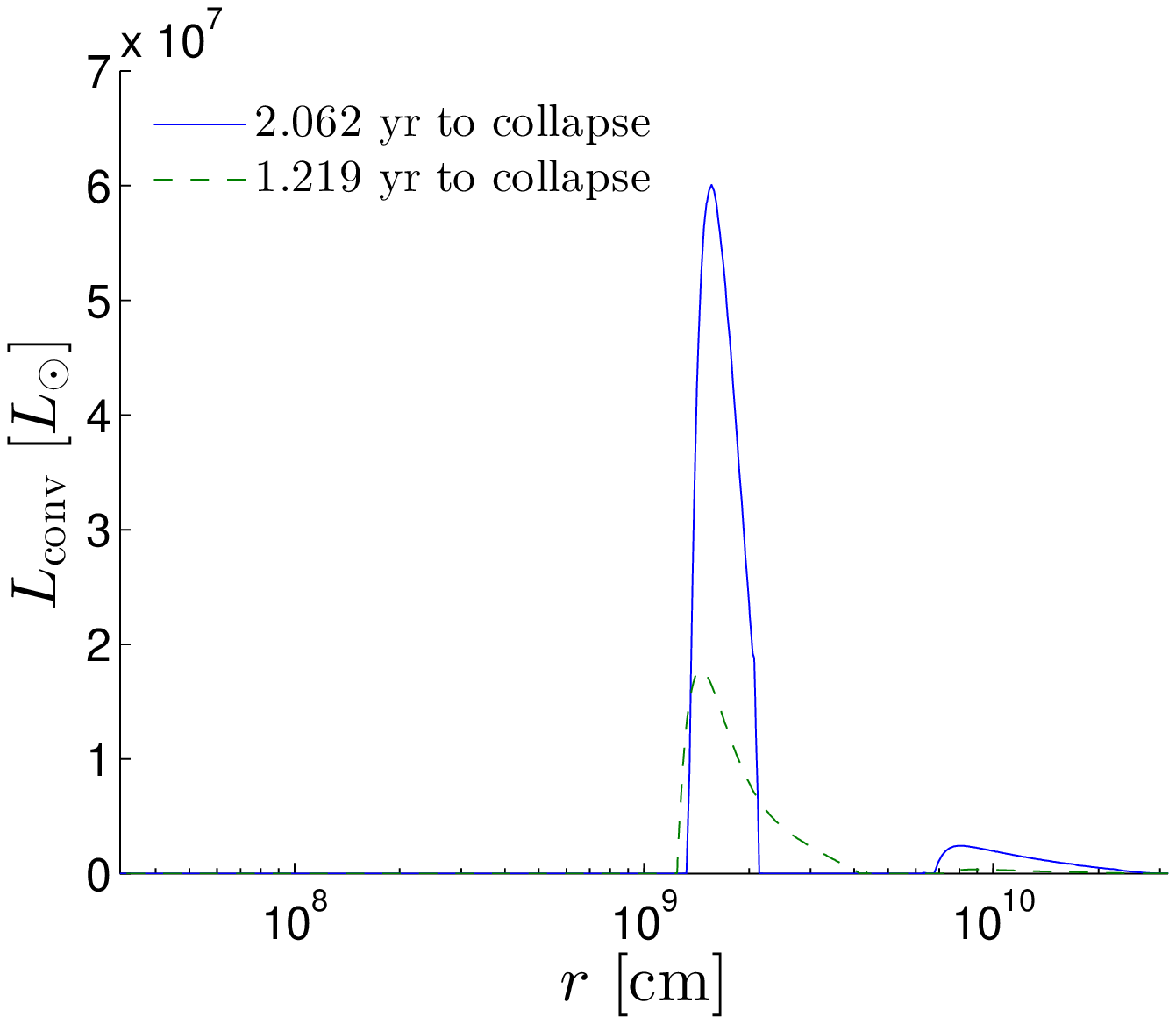}} &
	{\includegraphics*[scale=0.34]{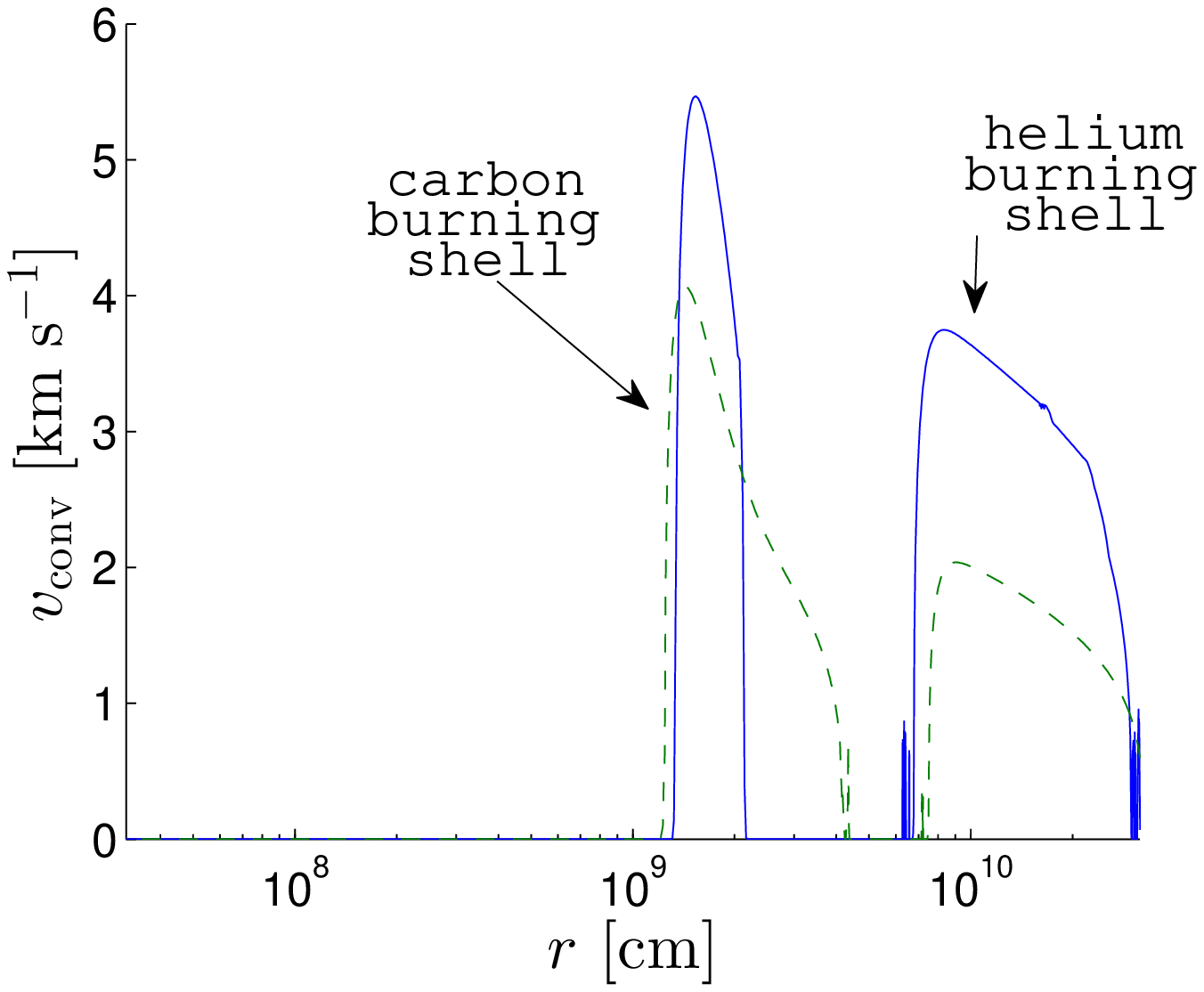}} &
	{\includegraphics*[scale=0.34]{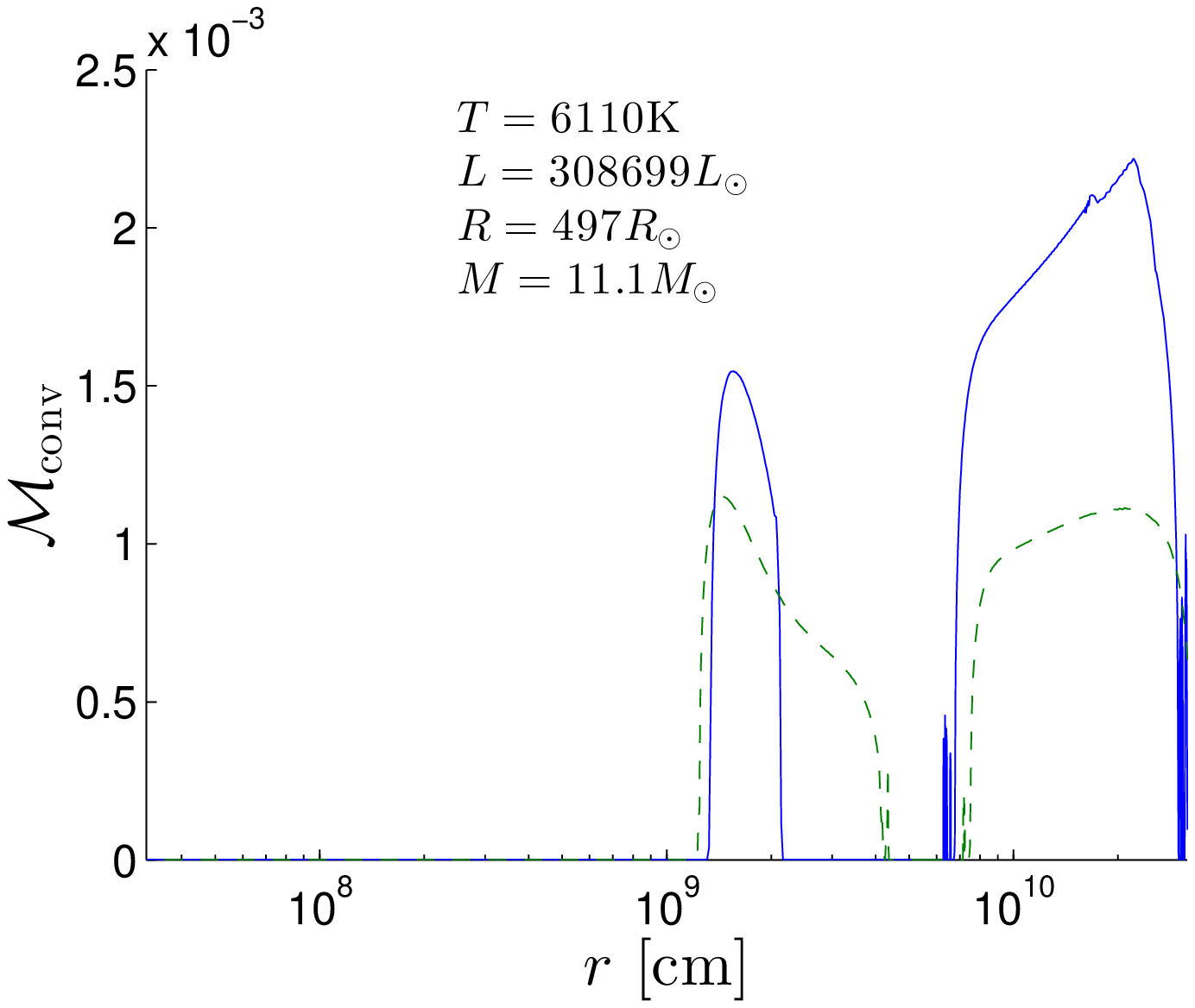}} \\
	{\includegraphics*[scale=0.34]{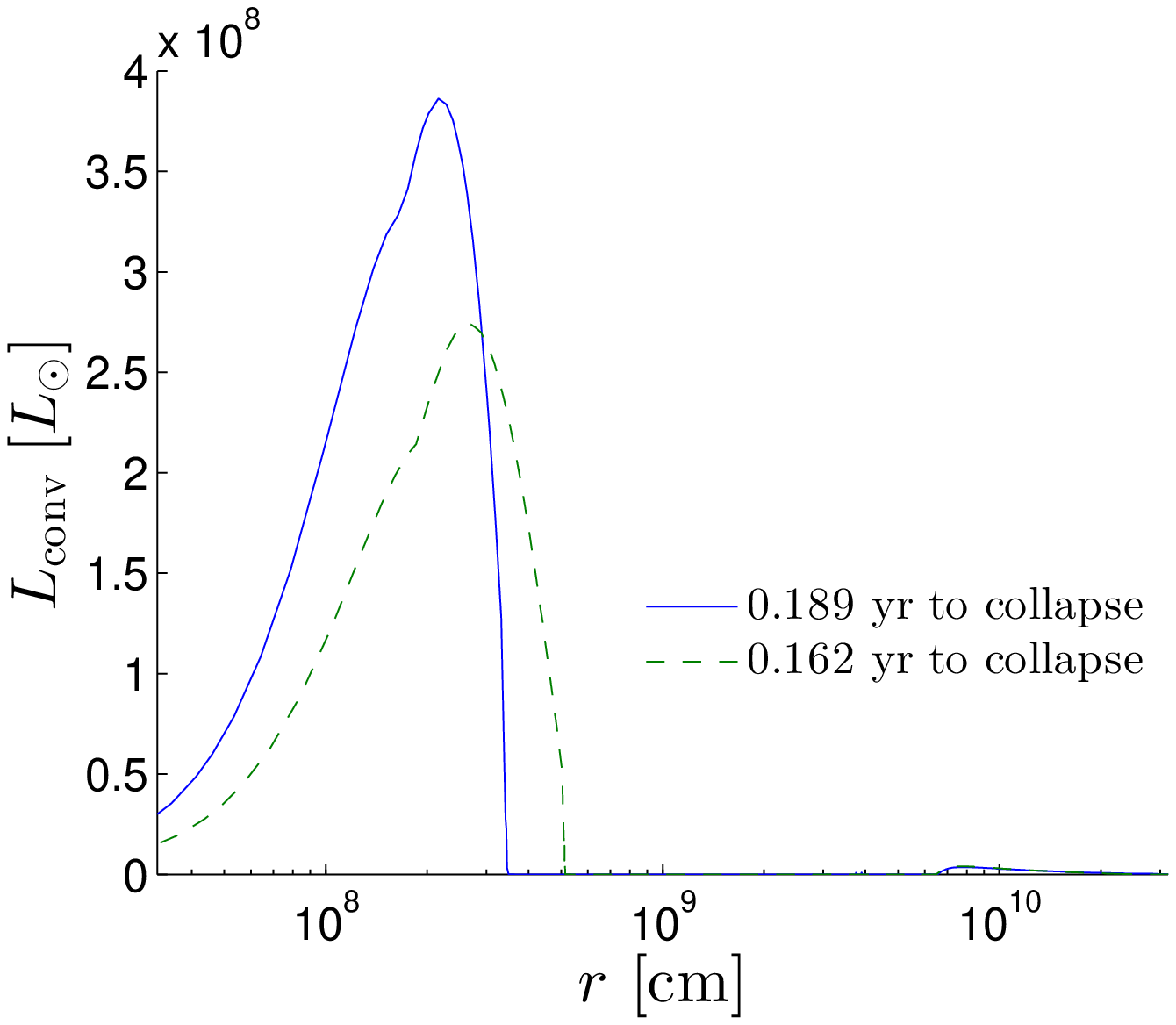}} &
	{\includegraphics*[scale=0.34]{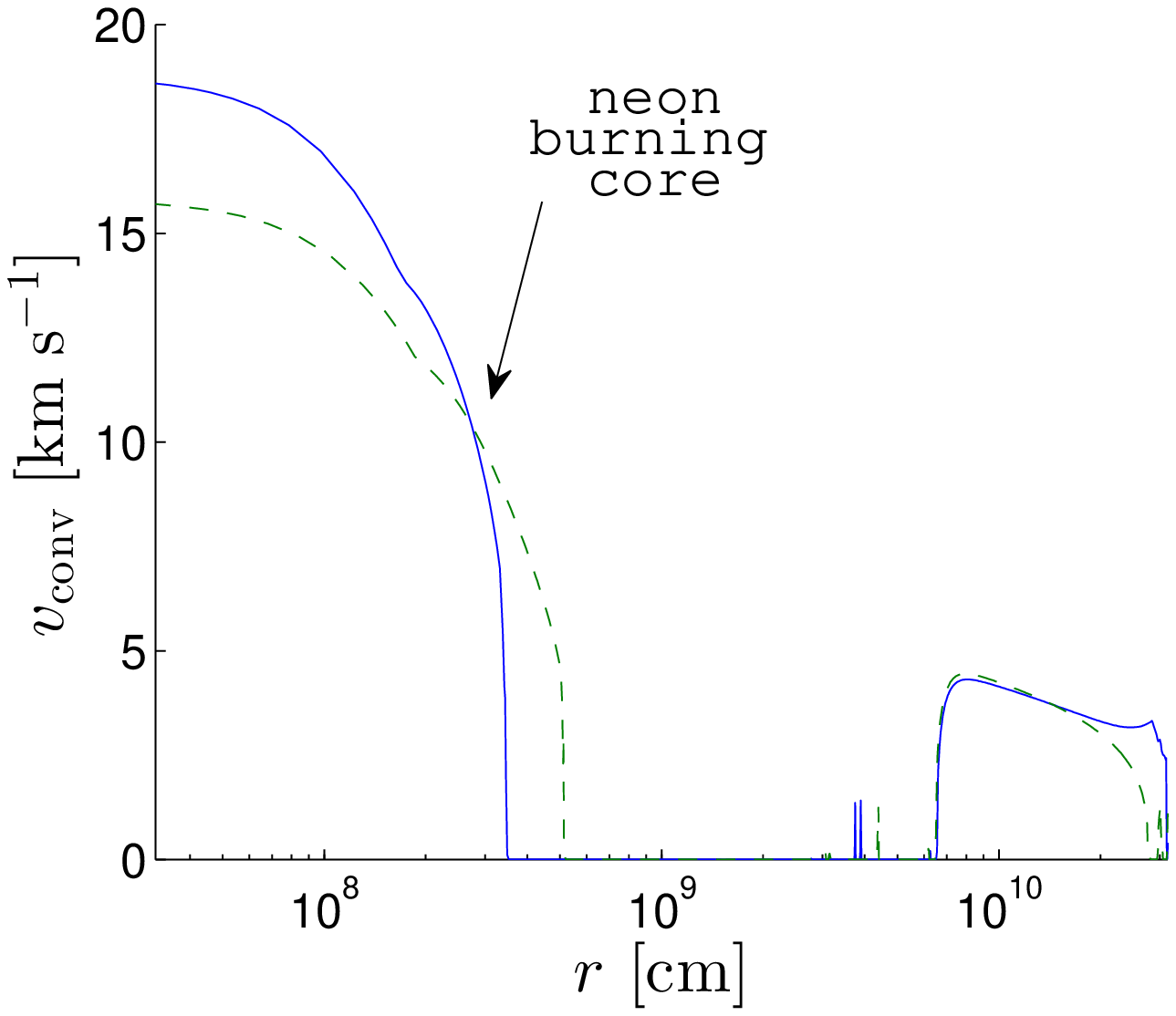}} &
	{\includegraphics*[scale=0.34]{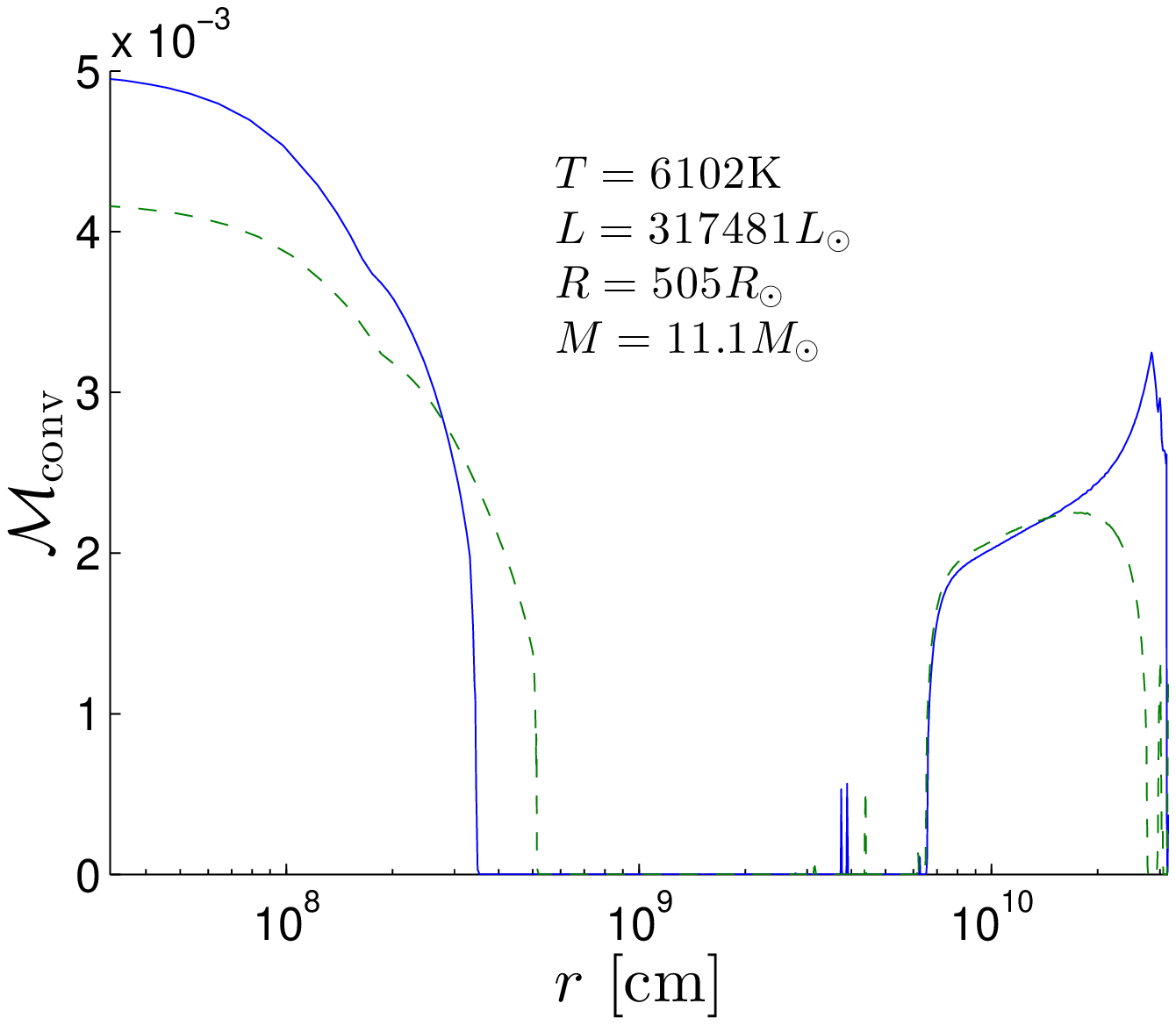}} \\
	{\includegraphics*[scale=0.34]{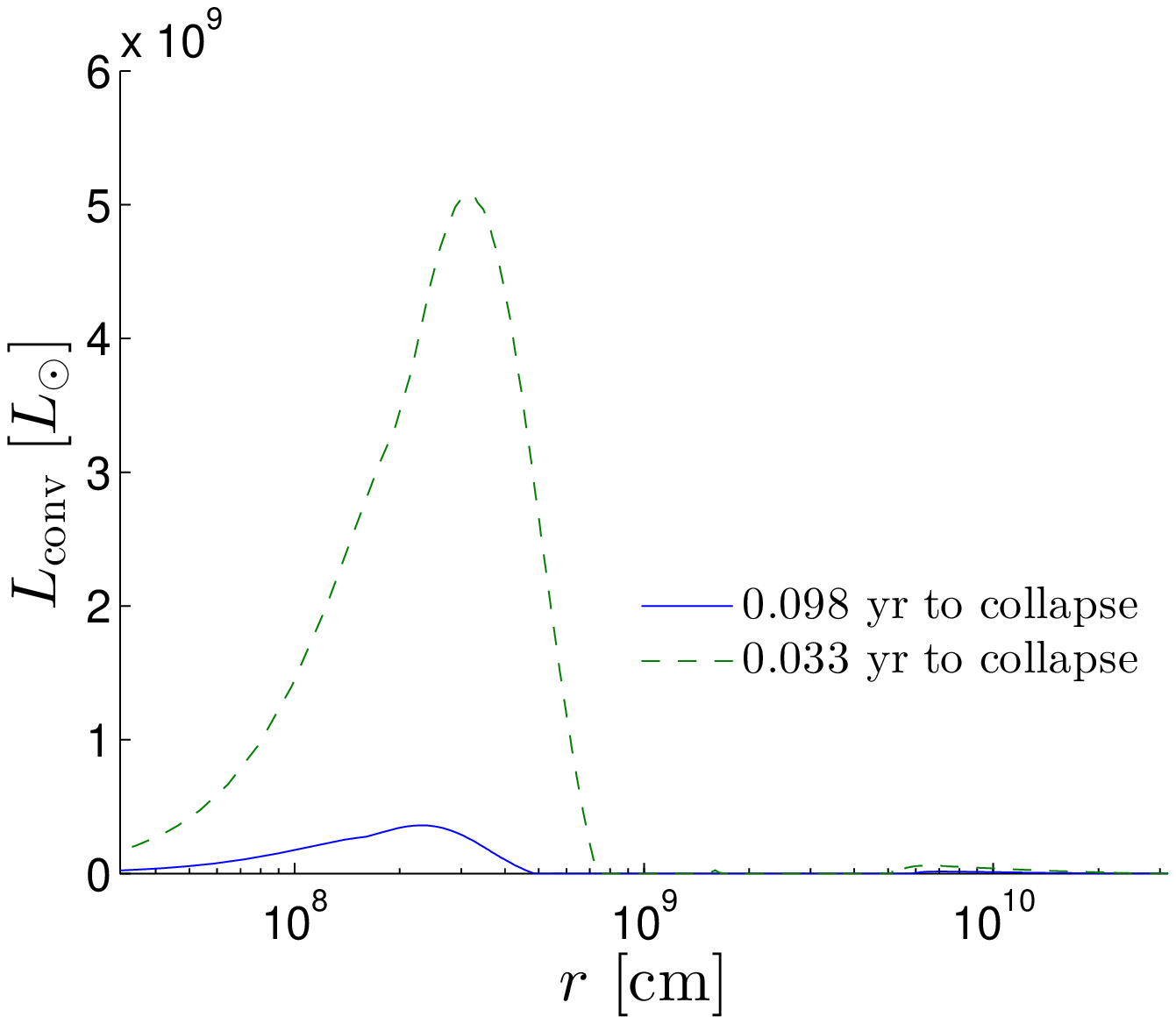}} &
	{\includegraphics*[scale=0.34]{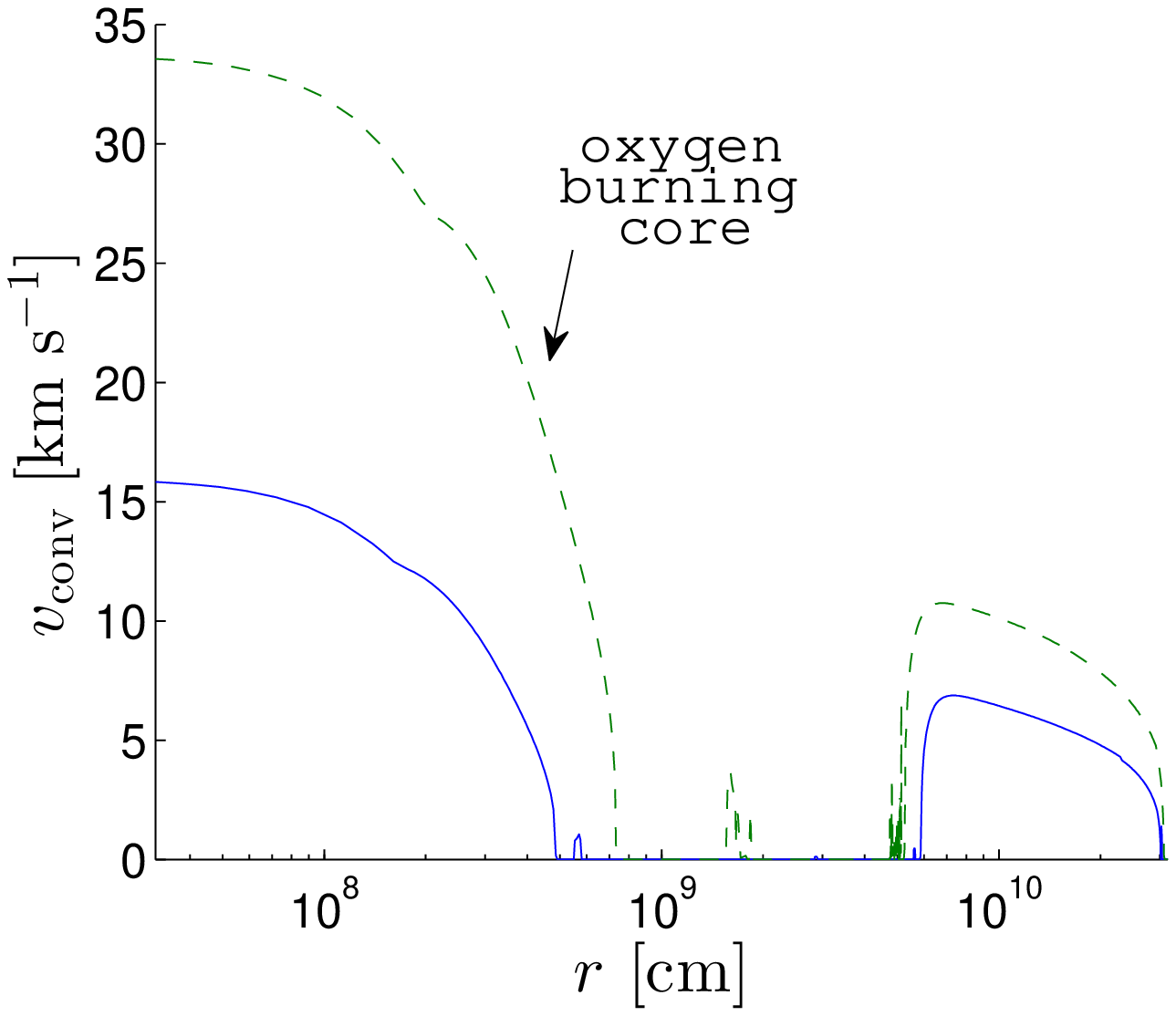}} &
	{\includegraphics*[scale=0.34]{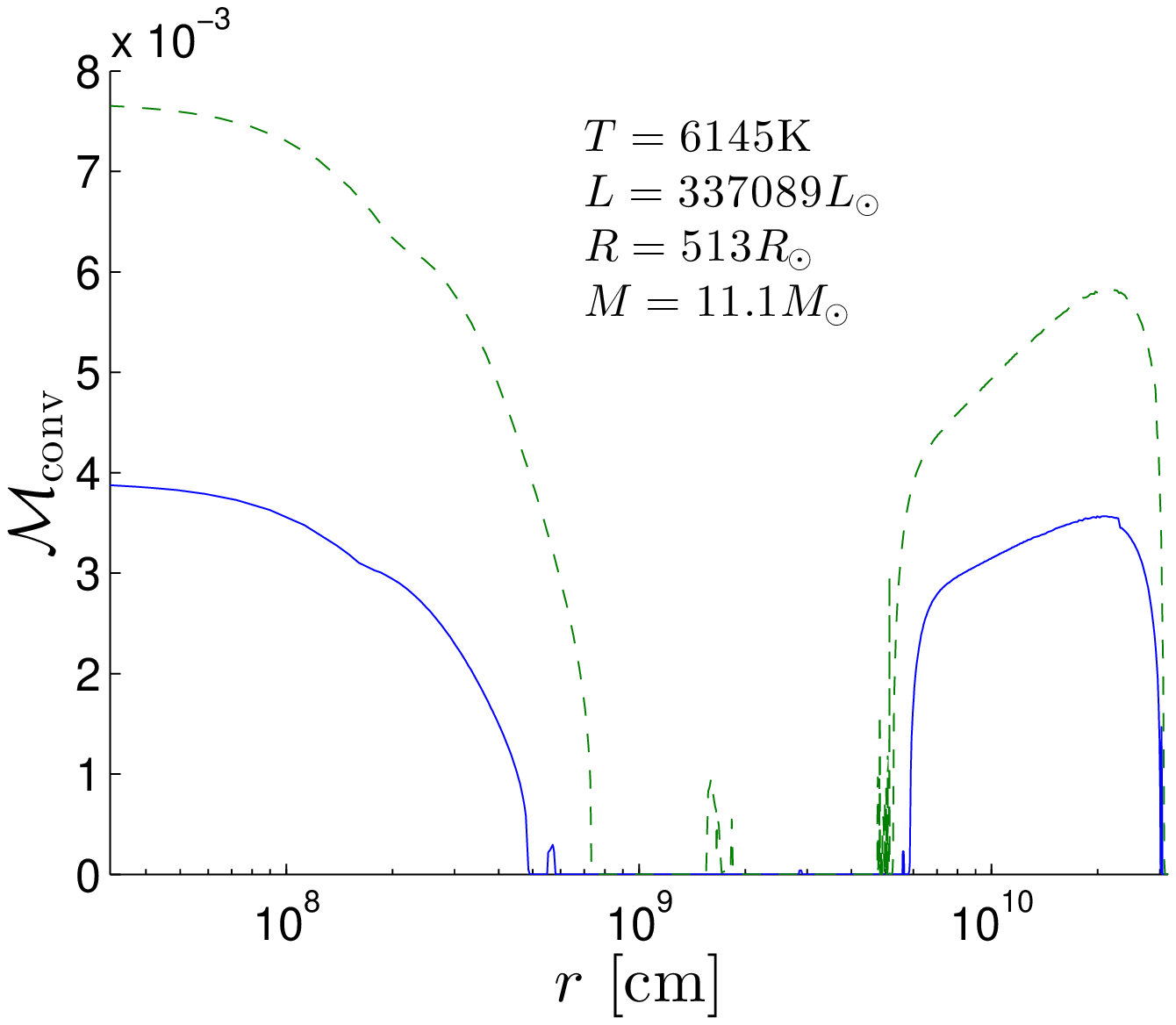}} \\
	{\includegraphics*[scale=0.34]{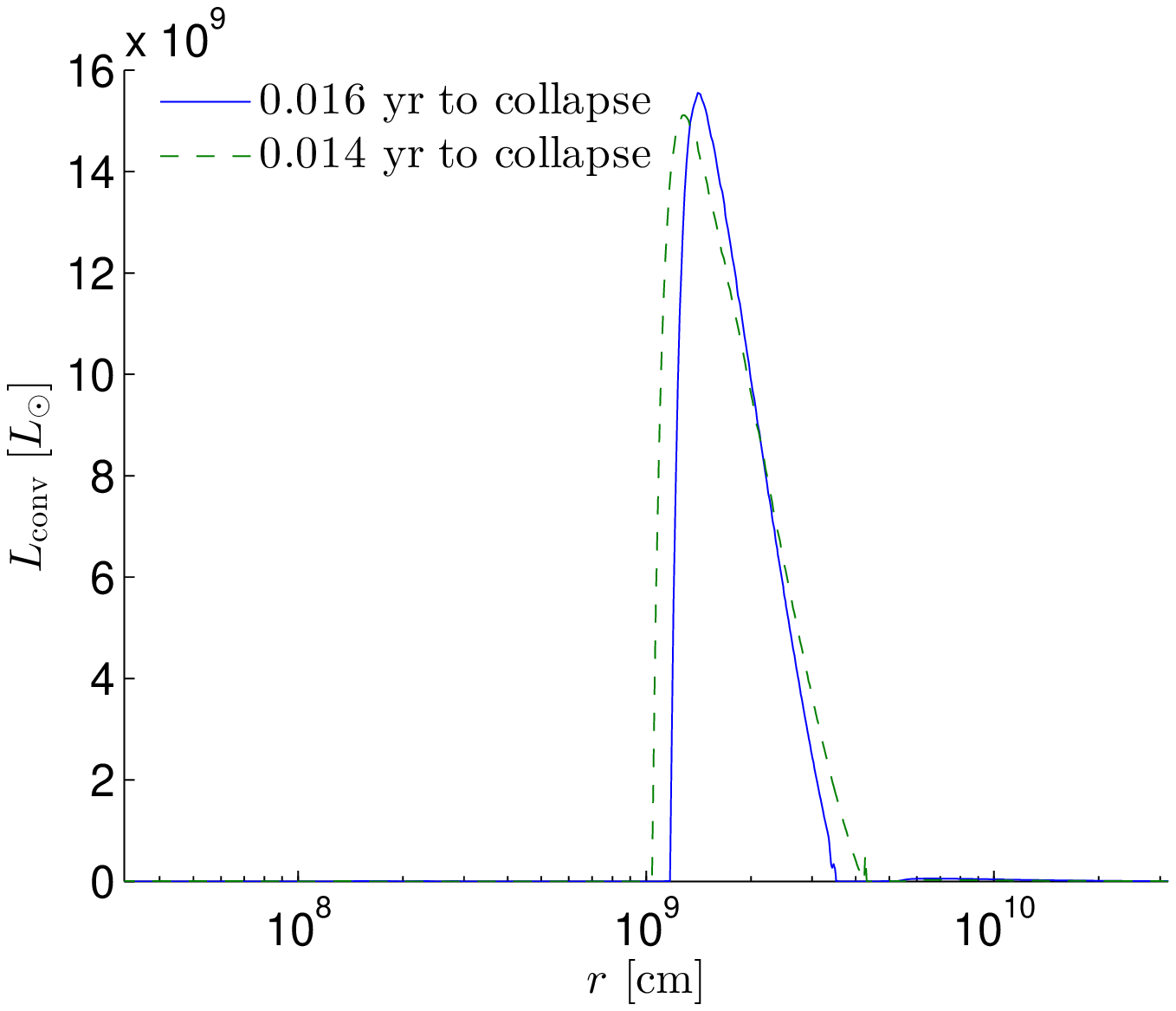}} &
	{\includegraphics*[scale=0.34]{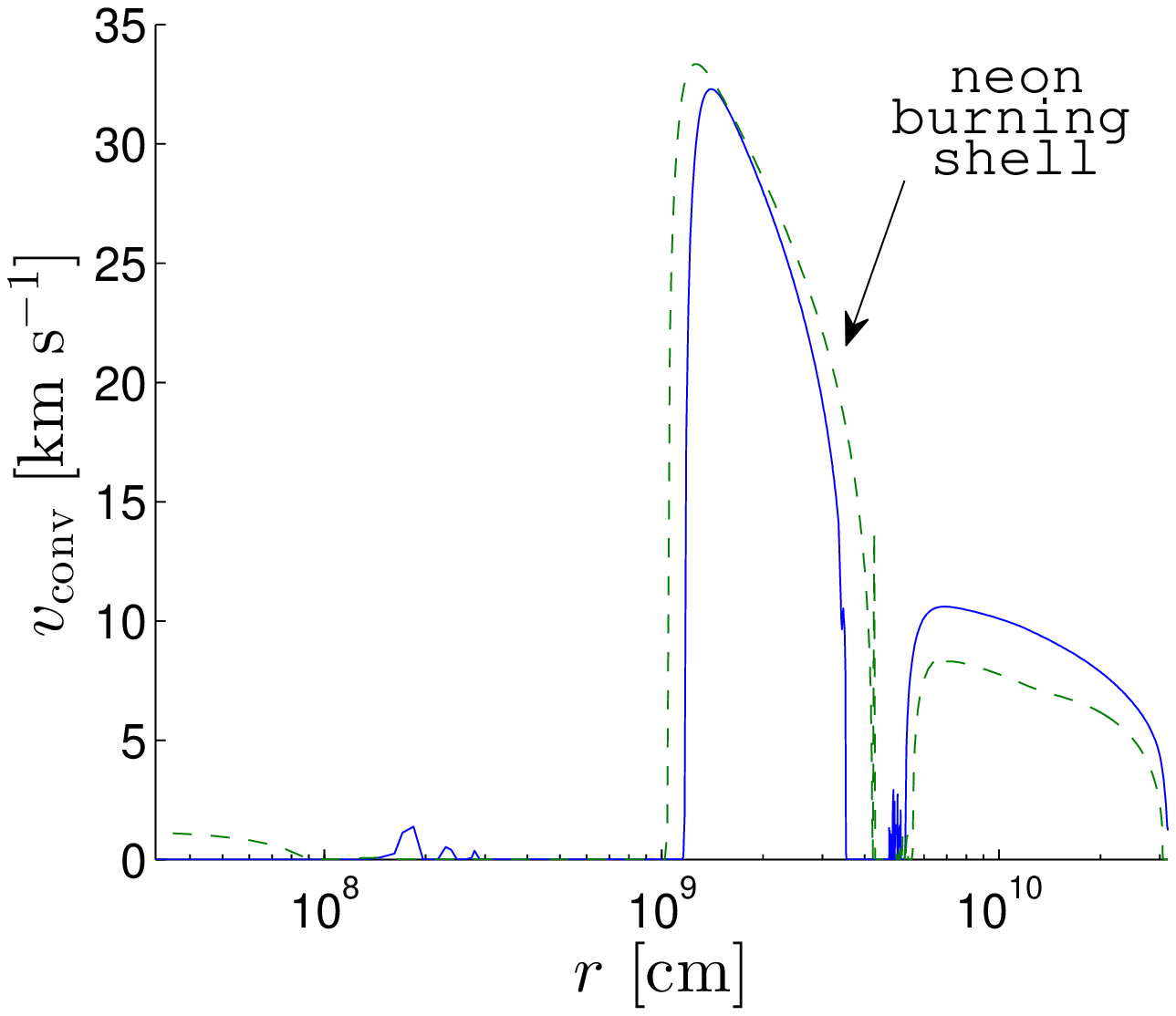}} &
	{\includegraphics*[scale=0.34]{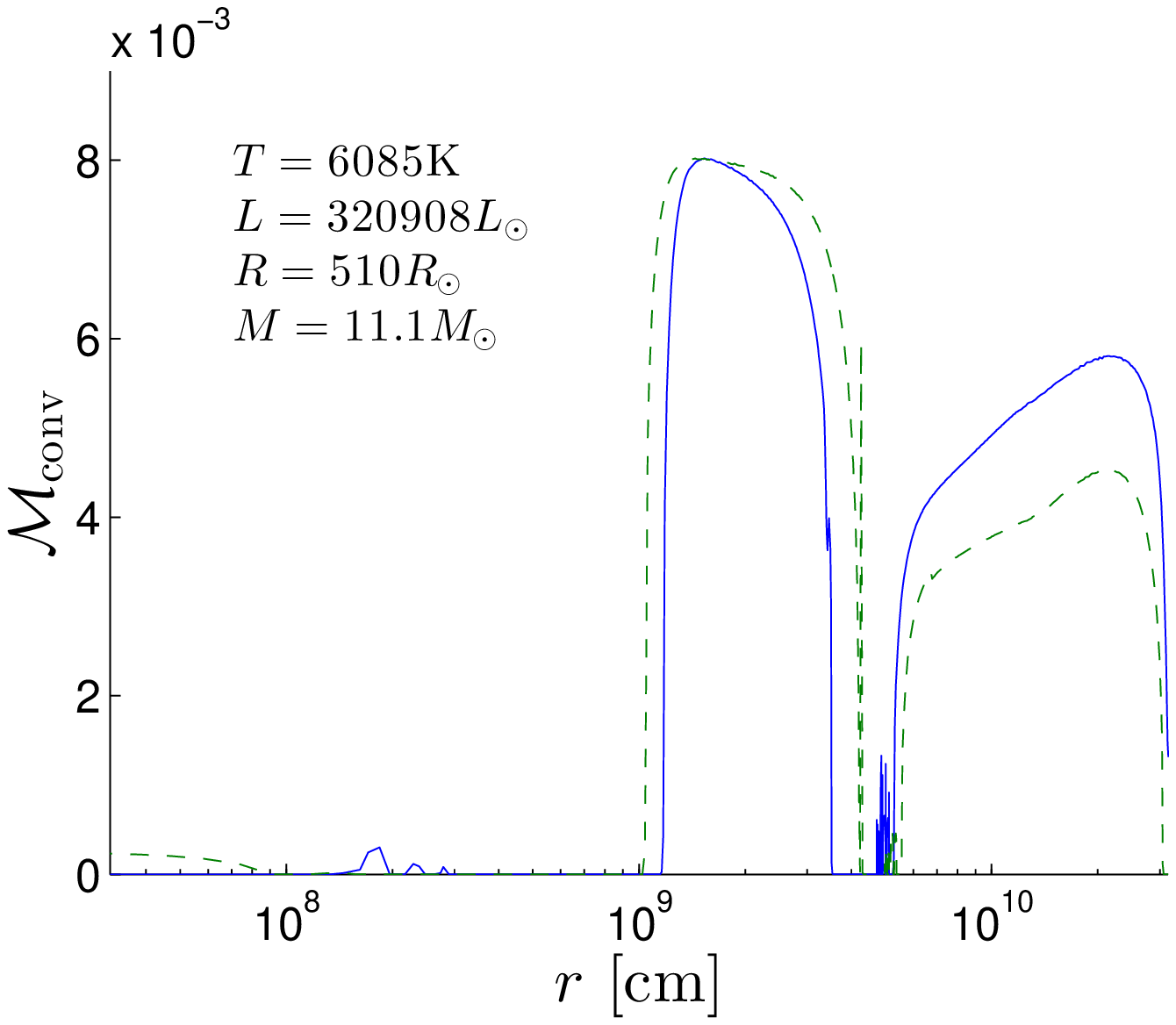}} \\
	{\includegraphics*[scale=0.34]{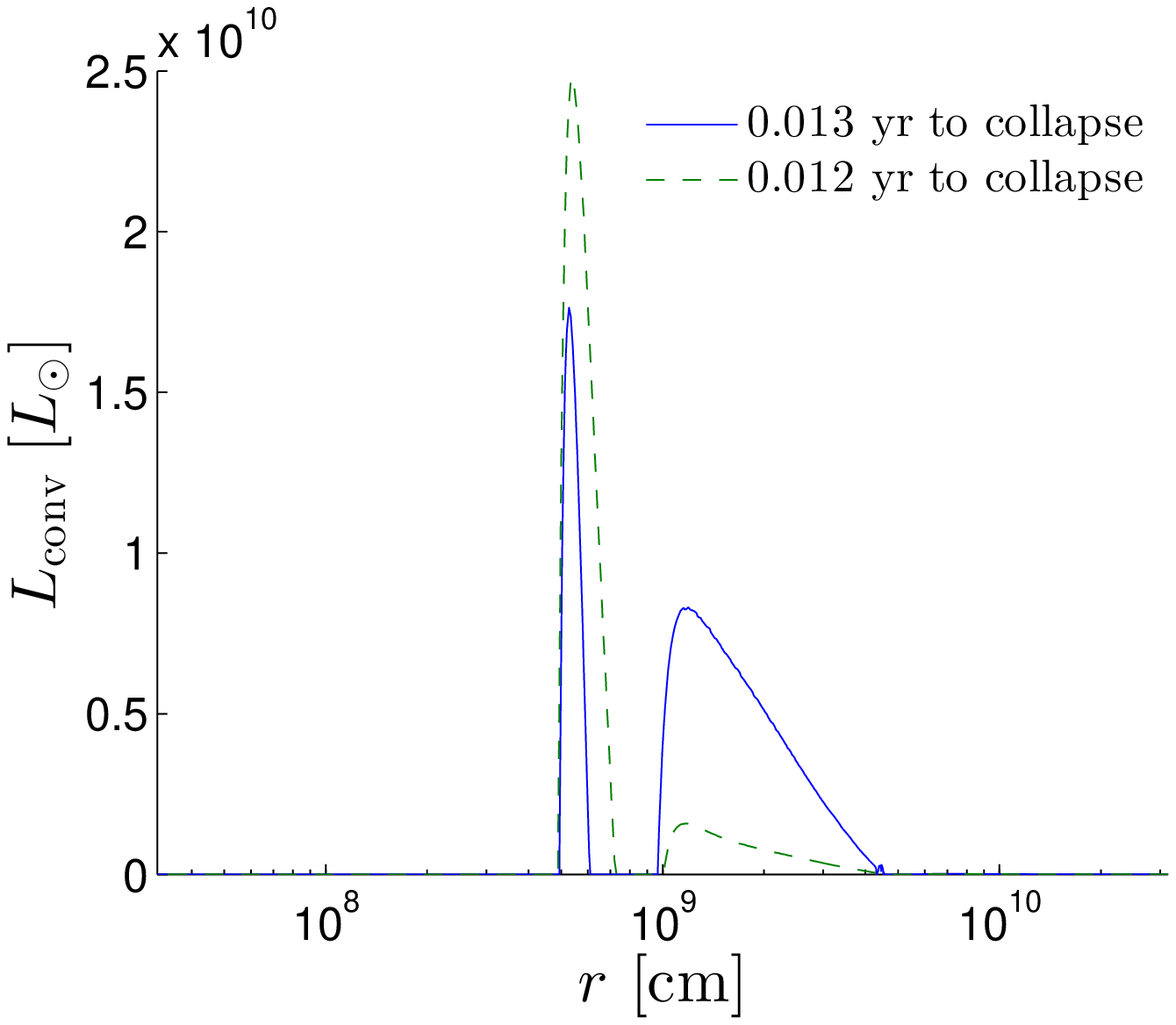}} &
	{\includegraphics*[scale=0.34]{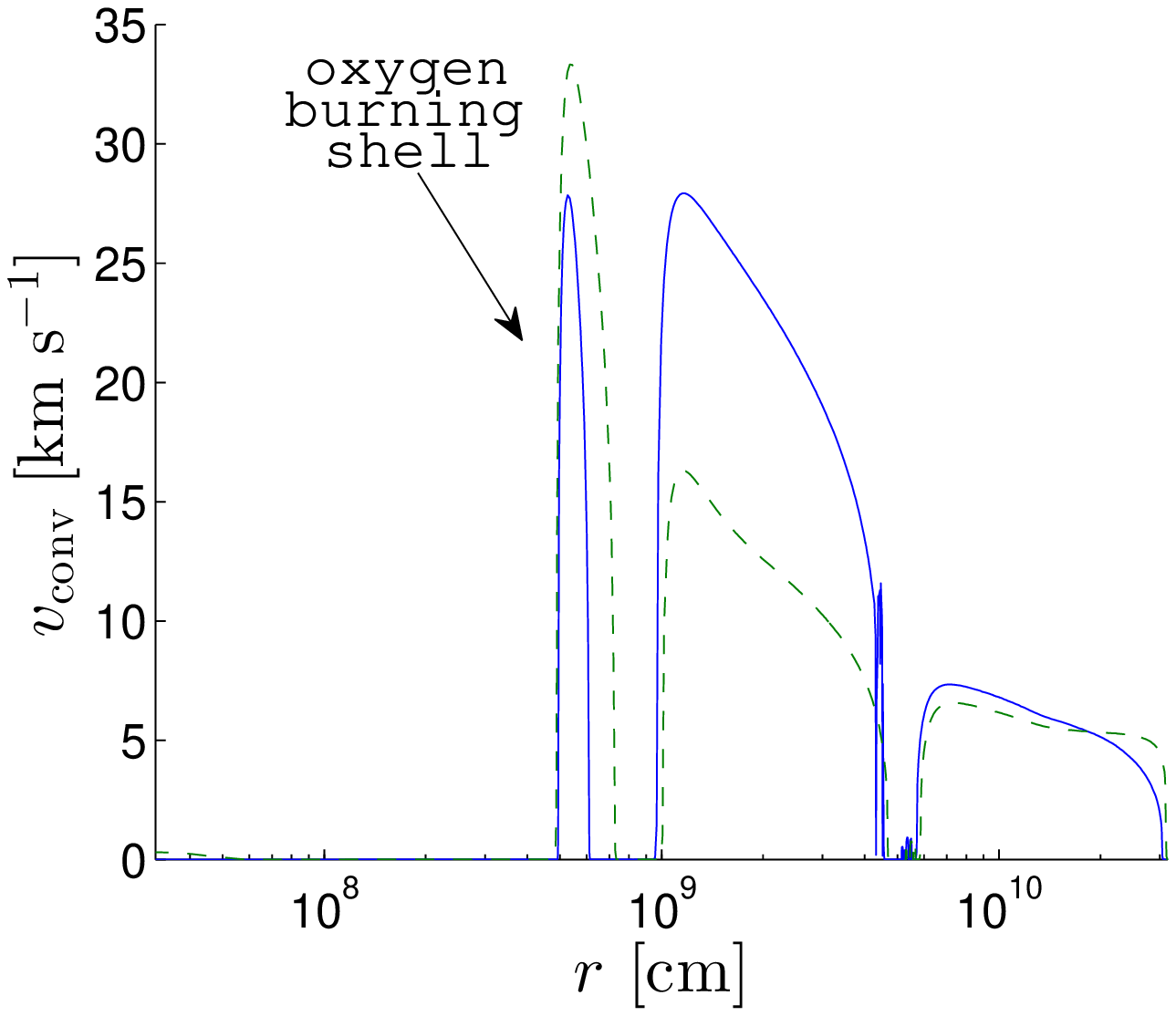}} &
	{\includegraphics*[scale=0.34]{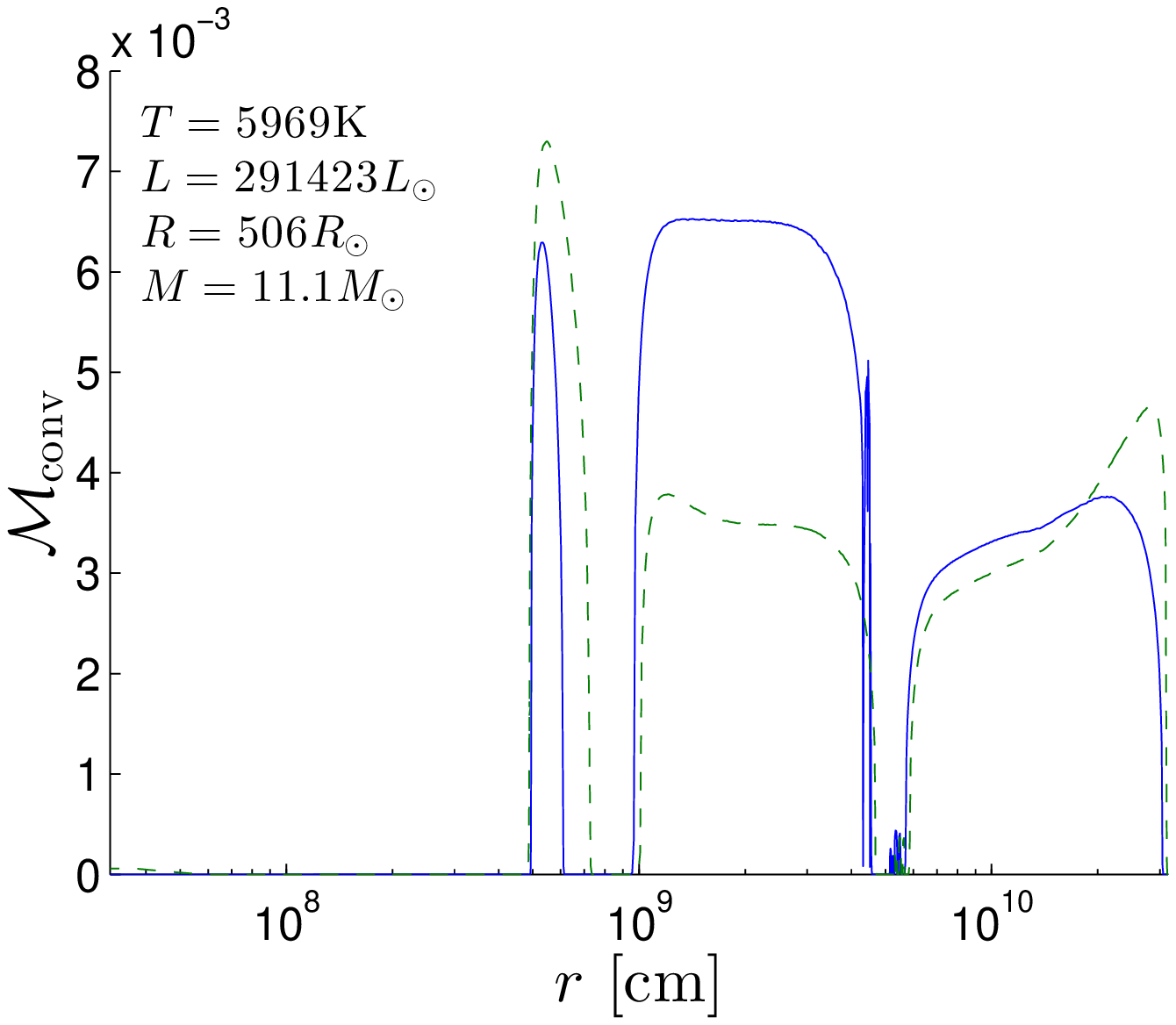}} \\
\else
	{\includegraphics*[scale=0.35]{M25LumCShell.eps}} &
	{\includegraphics*[scale=0.35]{M25VelCShell.eps}} &
	{\includegraphics*[scale=0.35]{M25MachCShell.eps}} \\
	{\includegraphics*[scale=0.35]{M25LumNeCore.eps}} &
	{\includegraphics*[scale=0.35]{M25VelNeCore.eps}} &
	{\includegraphics*[scale=0.35]{M25MachNeCore.eps}} \\
	{\includegraphics*[scale=0.35]{M25LumOCore.eps}} &
	{\includegraphics*[scale=0.35]{M25VelOCore.eps}} &
	{\includegraphics*[scale=0.35]{M25MachOCore.eps}} \\
	{\includegraphics*[scale=0.35]{M25LumNeShell.eps}} &
	{\includegraphics*[scale=0.35]{M25VelNeShell.eps}} &
	{\includegraphics*[scale=0.35]{M25MachNeShell.eps}} \\
	{\includegraphics*[scale=0.35]{M25LumOShell.eps}} &
	{\includegraphics*[scale=0.35]{M25VelOShell.eps}} &
	{\includegraphics*[scale=0.35]{M25MachOShell.eps}} \\
\fi
\end{tabular}
      \caption{Like Fig. \ref{fig:M15Conv} but for a star with $M_\zams = 25 M_\odot$.}
      \label{fig:M25Conv}
\end{figure*}

Most important for the dynamo activity is the core rotation profile and the convective zones. The rotation serves as the parameters that determines whether a magnetic activity takes place or not. This is the ingredient that makes the PEO a rare phenomena.
The basic assumption is that rapid rotation that leads to strong dynamo activity is caused by a stellar companion that merges with the core. This is a rare event. We here do not model this spin-up process of the core. The residual rotation in the simulated models is of no importance here because it is very small.
We concentrate here on the convective zones. Most significant are the velocities and Mach numbers of the convection cells, that are given according to the Mixing-Length Theory. These are presented in Figure \ref{fig:M15Conv} and Figure \ref{fig:M25Conv} for models with ZAMS masses of
$15 M_\odot$ and $25 M_\odot$, respectively, and at several stages of evolution.
Also presented is the convective luminosity at different stages.
The convective luminosity is defined as
\begin{equation}
L_\mathrm{conv} \equiv 4 \pi r^2 \rho v_\mathrm{conv}^3,
\label{eq:ConvLum}
\end{equation}
where $r$ is the shell radius,
$\rho$ is the gas density,
and $v_\mathrm{conv}$ is the convective velocity according to the Mixing-Length Theory.


The efficiency of the dynamo as operates in main sequence stars increases with decreasing value of the Rossby number, defined as the ratio of rotation period, $P_{\rm rot}$, to the convective overturn time, $\tau_c$, which is the mixing length divided by the convective velocity. To have a strong magnetic activity from dynamo amplification of magnetic fields, the Rossby number should be smaller than 1, and preferentially to be ${\rm Ro} \equiv (P_{\rm rot}/\tau_c) \la 0.1$ (e.g., \citealt{Pizzolatoetal2003}).
In our stellar models the Rossby number is ${\rm Ro} \approx 10$ in the regions of interest,
so that the core rotation rate needs to be higher by a factor of $10-100$ for an efficient dynamo.
Such fast pre-collapse rotations require a stellar binary companion to spin-up the core, most likely through a merger of the companion with the core.
In the present study we focus on the convective luminosity of inner shells,
and the study of the conditions for getting the required core rotation is deferred to future works.

\section{STELLAR EXPANSION}
\label{sec:tubes}

In this section we try to estimate the influence of the buoyantly-rising magnetic flux tubes on the envelope.
There are two kinds of drag force that act on a rising flux tube.
The first one acts only in convective regions and is due to the force exerted by the motion of convective cells. It imposes a minimum value on the magnetic field for the flux tube to move upward (e.g.,  \citealt{Fan2009b}; eq. 29 there). Here we assume that the magnetic flux tube obeys this condition, and rises up at a velocity of the order of magnitude of the convective velocity (as in the sun, e.g., \citealt{Fan2009a}). The typical velocity of convective cells is $v_{\rm conv}(r) \ga 0.001 c_s(r)$, where $c_s(r)$ is the local sound speed. In the inner burning shells, carbon burning and inward, $v_{\rm conv} \ga 3 \km \s^{-1}$ (see  Fig. \ref{fig:M15Conv} and Fig. \ref{fig:M25Conv}). This implies that the flux tube crosses the convective shell where it forms inside the core in a time of $t_1 \approx r_1/v_{\rm conv} \approx 1$~hour.
This time is very short compared with the pre-explosion time of mass loss, and can be neglected.

We assume that after the flux tube exits the convective shell in the core, where it has been formed, its velocity rises much above that of the convective velocity. The terminal velocity of the flux tube is derived from balancing the buoyancy force with the drag force acting on the rising flux tube.
The terminal velocity of a cylindrical flux tube rising with its long axis being horizontal
is given by (e.g., \citealt{Parker1975})
\begin{equation}
v_{\rm tube} \simeq \left( \frac{P_{B}}{P_\ast} \right)^{1/2}
\left( \frac{a}{H_P} \right)^{1/2}
\left( \frac{\pi}{\gamma C_D} \right)^{1/2} c_s ,
\label{eq:vtube}
\end{equation}
where $a$ is the radius of the cross section of the flux tube, $P_B$ is the magnetic pressure inside the flux tube, $P_\ast(r)$ is the pressure in the star at the location of the tube, $C_D \approx 1$ is the drag coefficient, $\gamma$ is the adiabatic index, and $H_P$ is the pressure scale height.

The magnetic energy generated in the core is assumed to be transported by many flux tubes, much as the process in the sun is.
Consider several flux tubes crossing radius $r$ at a speed $v_{\rm tube}$ and covering an area $\eta 4 \pi r^2$ on the spherical sphere of radius $r$. The magnetic power is given by
\begin{eqnarray}
L_B & \approx & \eta P_B 4 \pi r^2  v_{\rm tube} \nonumber \\
& \approx &
\eta
\left( \frac{P_B}{P_\ast} \right)^{3/2}
\left( \frac{a}{H_P} \right)^{1/2}
\left( \frac{\pi}{\gamma^3 C_D} \right)^{1/2}  L_{\rm max,conv},
\label{eq:Lb}
\end{eqnarray}
where
in the second equality we multiply by $\rho c_s^2/(\gamma P_\ast)=1$, and
\begin{equation}
L_{\rm max,conv} = 4 \pi \rho r^2 c_s^3,
\label{eq:lmaxconv1}
\end{equation}
is the maximum power that subsonic convection can carry \citep{QuataertShiode2012}.
The radius $r_{d}$ (marked $r_{\rm ss}$ by  \citealt{ShiodeQuataert2014}) is where
subsonic convection is unable to carry the outgoing wave energy, i.e
$L_{\rm wave} > L_{\rm max,conv} (r_{d})$. According to \cite{QuataertShiode2012}, the inability of subsonic convection to carry the outgoing wave energy implies that the wave power will inevitably drive an outflow.
The quantity $L_{\rm max,conv} (r)$ decreases very steeply with radius \citep{ShiodeQuataert2014}. Therefore, even if the magnetic power is smaller than the waves-power considered by \cite{ShiodeQuataert2014}, $L_B \approx 0.1 L_{\rm wave}$, we have at the radius $r_d$ considered by them
\begin{eqnarray}
\left( \frac{P_B}{P_\ast} \right)^{3/2}   \approx  0.1  \eta^{-1}
\frac{L_B}{0.1 L_{\rm wave}}
\left( \frac{H_P}{a} \right)^{1/2}
\left( \frac{\gamma^3 C_D}{\pi} \right)^{1/2}.
\label{eq:Pb}
\end{eqnarray}

Since $\eta<1$, and likely $\eta<0.5$, $a \la H_P$,
and with $C_D=1$ and $\gamma=5/3$ the last term in equation (\ref{eq:Pb}) is 1.2, the right hand side of equation (\ref{eq:Pb}) can be written as $\approx 1 (L_B/0.1 L_{\rm wave})$.
Taking $L_B \approx 0.1 L_{\rm wave}$ (section \ref{sec:dynamo}), we find that the right hand side of equation (\ref{eq:Pb}) is about equal to 1 in the outer region where the waves in the model of \cite{QuataertShiode2012} and \cite{ShiodeQuataert2014} are strongly dissipated and assumed to eject mass. Hence, the magnetic pressure in the proposed model and in a case of a magnetically active core obeys $P_B (r_d) \approx P_\ast(r_d)$. This implies that the buoyant velocity of the flux tubes is about the sound speed, as the waves. Energy deposition can lead to envelope inflation \citep{McleySoker2014}.

We summarize this section by stating that under our assumptions, a magnetically active pre-explosion core might significantly perturb the envelope and lead to its expansion. The rising time of the flux tubes in case of a powerful dynamo in the core is only slightly longer than the propagation time of waves from the core to the outer envelope. The motion of the flux tubes and the volume they occupied increases the pressure in the envelope. Dissipation of magnetic energy, e.g., by reconnection, will increase the thermal pressure. The increase in pressure,
both directly from the flux tubes and from thermal energy will cause the star to expand.
As previously discussed, the expansion will lead to enhanced mass loss rate as well as to a possible interaction with a binary companion.

\section{SUMMARY}
\label{sec:summary}

We addressed the still open question of the reasons for the occurrence of PEOs in about one in ten CCSNe. The burning shells have phases with very high convective luminosity, starting with the carbon burning shells tens of years before explosion, and ending with the silicon burning shell.
Figures \ref{fig:M15Conv} and \ref{fig:M25Conv} emphasize the burning shells of carbon, neon, and oxygen. If the strong convection is coupled with sufficiently rapid core rotation (section \ref{sec:stellar}), strong magnetic activity might take place (section \ref{sec:dynamo}). We speculated that such an activity can result in magnetic flux tubes that rise to the envelope, much as is the case in the sun, and more so in magnetically active main sequence stars. The magnetic energy deposited in the outer envelope (section \ref{sec:tubes}) might lead to effects similar to those proposed for waves exited by core convection \citep{QuataertShiode2012, ShiodeQuataert2014}.

The main effect of the energy deposition to the envelope is likely to be envelope expansion \citep{McleySoker2014} and enhanced mass loss rate. The enhanced mass loss rate will play a role during carbon shell burning that lasts relatively for a long time (tens of years in not too-massive stars). More violent processes will take place if a close binary companion accretes mass from the expanding envelope.
The accretion process onto the secondary star releases more energy. In particular the secondary star might launch jets.

\cite{HarpazSoker2009} suggested a scenario of a magnetic activity, but in the envelope rather than in the core, for the 1837-1856 Great Eruption of the LBV $\eta$ Carinae. Most likely, the secondary star in $\eta$ Carinae accreted mass and launched two jets that shaped the bipolar nebula, the Homunculus, around $\eta$ Carinae \citep{KashiSoker2010}. The speculative scenario we have proposed makes a connection between PEOs and major eruptions of LBV.

The requirement for some minimum rotation velocity (section \ref{sec:stellar}) for a strong  dynamo activity and the stochastic nature of magnetic activity, account for the rare occurrence of PEOs in the proposed scenario.

As the core eventually collapses, strong magnetic fields might influence the explosion mechanism. The core rotation and strong magnetic fields will be amplified in the gas around the newly born neutron star. Rapid rotation and magnetic fields are thought to facilitate the formation of jets around compact objects. Such jets might explode the star \citep{PapishSoker2011, GilkisSoker2014}, and might even lead to a very energetic CCSN \citep{Gilkisetal2016}.

\section*{Acknowledgments}

We thank an anonymous referee for very helpful comments.

\ifmnras
	\bibliographystyle{mnras}

\label{lastpage}

\end{document}